\documentclass[preprint]{revtex4}
\usepackage{graphicx}
\usepackage{amsmath}
\usepackage{epsfig}
\usepackage{amsfonts, color, soul}

\newcommand{\X}{${\rm X}$}
\newcommand{\Xa}{${\rm X^*}$}
\newcommand{\Ea}{${\rm E_a}$}
\newcommand{\EaX}{${\rm E_aX} $}
\newcommand{\Ed}{${\rm E_d}$}
\newcommand{\EdXa}{${\rm E_dX^*}$}

\newcommand{\cX}{{\rm [X]}}
\newcommand{\cXT}{{\rm [X]_T}}
\newcommand{\cST}{{\rm [S]_T}}
\newcommand{\cXa}{{\rm [X^*]}}
\newcommand{\cXaT}{{\rm [X^*]_T}}
\newcommand{\cEa}{{\rm [E_a]}}
\newcommand{\cEaX}{{\rm [E_aX]}}
\newcommand{\cEaT}{{\rm [E_a]_T}}
\newcommand{\KM}{K_{\rm M}}
\newcommand{\KMa}{K_{\rm M,a}}
\newcommand{\KMd}{K_{\rm M,d}}
\newcommand{\cEd}{{\rm [E_d]}}
\newcommand{\cEdXa}{{\rm [E_dX^*]}}
\newcommand{\cEdT}{{\rm [E_d]_T}}

\begin{document}

\title{Enzyme localization can drastically affect signal amplification
  in signal transduction pathways}
\author{Siebe B. van Albada}
\affiliation{FOM Institute for Atomic and Molecular Physics,
Kruislaan 407, 1098 SJ Amsterdam, The Netherlands.}

\author{Pieter Rein ten Wolde}
\affiliation{FOM Institute for Atomic and Molecular Physics,
Kruislaan 407, 1098 SJ Amsterdam, The Netherlands.}

\begin{abstract}
  Push-pull networks are ubiquitous in signal transduction pathways in
  both prokaryotic and eukaryotic cells. They allow cells to strongly
  amplify signals via the mechanism of zero-order ultrasensitivity. In
  a push-pull network, two antagonistic enzymes control the activity
  of a protein by covalent modification.  These enzymes are often
  uniformly distributed in the cytoplasm. They can, however, also be
  colocalized in space, for instance, near the pole of the
  cell. Moreover, it is increasingly recognized that these enzymes can
  also be spatially separated, leading to gradients of the active form
  of the messenger protein. Here, we investigate the consequences of
  the spatial distributions of the enzymes for the amplification
  properties of push-pull networks. Our calculations reveal that
  enzyme localization by itself can have a dramatic effect on the
  gain.  The gain is maximized when the two enzymes are either
  uniformly distributed or colocalized in one region in the
  cell. Depending on the diffusion constants, however, the sharpness
  of the response can be strongly reduced when the enzymes are
  spatially separated. We discuss how our predictions could be tested
  experimentally.
\end{abstract}

\maketitle

\noindent {\bf Synopsis}

\noindent Living cells continually have to respond to a changing
environment. To this end, they do not only have to detect
environmental signals, but also amplify them. In living cells, signals
are often amplified in so-called push-pull networks. In a push-pull
network, two enzymes control the activity of a protein in an
antagonistic manner. A well-known example is a network in which a
kinase phosphorylates a messenger protein, while a phosphatase
dephosphorylates the same protein. While it has long been assumed
that the enzymes are uniformly distributed in the cytoplasm, it is
increasingly becoming clear that in many systems one or both of the enzymes are
localized in space, for instance near the cell pole. If the enzymes
are spatially separated, then spatial gradients of the messenger
protein can form, and recently a number of these protein gradients
have been observed experimentally. We study by
numerical calculations how the amplification properties of push-pull
networks depend upon the spatial distribution of the enzymes. We
find that the gain is maximized when the enzymes are either uniformly
distributed or colocalized in space. Depending upon the diffusion
constants, however, the sharpness of the response can be strongly
reduced when the enzymes are spatially separated. 

\newpage

\noindent {\bf \sf \Large Introduction}\\[0.2cm]
Living cells are information processing machines.  In order to process
information reliably, signals often need to be amplified. To this end,
cells can employ a variety of amplification mechanisms. Signals can be
amplified via positive feedback, cooperative binding of signaling
molecules to receptors, or interactions between receptor
molecules~\cite{Bray98}. Another principal mechanism for signal
amplification is zero-order
ultrasensitivity~\cite{Goldbeter81,Goldbeter84}. This mechanism 
operates in so-called push-pull networks, which are omnipresent in
both prokaryotes and eukaryotes. In a push-pull network, two enzymes
covalently modify a component in an antagonistic manner (see
Fig.~\ref{fig:fig1}). One well-known example is a network in which a
kinase phosphorylates a component, and a phosphatase dephosphorylates
the same component. If both
enzymes operate near saturation, then the modification reactions
become zero order, which means that the reaction rates become
insensitive to the substrate concentrations. Under these conditions, a
small change in the concentration of one of the two enzymes (the input
signal), will lead to a large change in the concentration of the
modified protein (the output
signal)~\cite{Goldbeter81,Goldbeter84}. The amplification properties
of push-pull networks have been analyzed in
detail~\cite{Goldbeter81,Goldbeter84,Ferrell96,Berg00,Detwiler00,Samoilov05,TanaseNicola06}.
In these studies, however, it is assumed that the antagonistic enzymes are
uniformly distributed in space. Yet, it is increasingly recognized
that in many systems one or both of the two antagonistic enzymes are localized
in space, for instance at the cell pole.  Here, we address the
question how the spatial distribution of the antagonistic enzymes
affects the amplification properties of push-pull networks.

If the two antagonistic enzymes are separated in space, then gradients
of the messenger protein can
form~\cite{Brown99_2,Vaknin04,Lipkow05,Rao05,Kholodenko06}. Recently,
a number of protein gradients have been observed experimentally in
both prokaryotic and eukaryotic cells. For example, in {\em
  Escherichia coli} cells, the kinase CheA and the phosphatase CheZ
control the phosphorylation level of the messenger CheY, which
transmits the chemotactic signal from the receptor cluster to the
flagellar motors. In wild-type cells, the kinase and the phosphatase
are both localized at the receptor cluster~\cite{Sourjik00}, and, as a
result, the steady-state concentration profile of CheY is
uniform~\cite{Vaknin04}. However, in {\em E. coli} mutants, where the
phosphatase is distributed in the cytoplasm, gradients of CheY have
recently been observed~\cite{Vaknin04}. Other examples of protein
gradients include {\em Caulobacter}, in which MipZ gradients guide
chromosome segregation and division site
selection~\cite{Thanbichler06}. In eukaryotic cells, gradients of Ran,
Stathmin, and HURP proteins aid in the formation of the mitotic
spindle by providing directional cues for microtubule
growth~\cite{Kalab02,Niethammer04,Caudron05,Wong06}. Moreover, in
eukaryotic cells, the kinases in the mitogen-activated protein kinase
(MAPK) cascade often bind to scaffold proteins, while the phosphatases
are distributed in the cytoplasm~\cite{Braybook}. This will lead to
concentration gradients of the activated kinases, which can become
particularly important if the scaffolds are located near the membrane.

In this study, we compare the amplification properties of a canonical
push-pull network, where all components are uniformly distributed in
space, with those of a network in which the enzyme that provides the
input signal is localized at one end of the cell, while all the other
components can freely diffuse through the cell. In the latter case,
the concentration profile of the messenger---the output signal---is
non-uniform. Previous studies have focused on the time-dependent
concentration profiles of the
messenger~\cite{Brown99_2,Rao05,Kholodenko06} and on the `control' of
diffusion over protein fluxes~\cite{Kholodenko00} in similar
systems. Here, we examine the effect of the spatial distribution of
the enzymes on the amplification properties of push-pull networks. To this end, we compute for both systems the
steady-state input-output relations. Our analysis reveals that the
spatial distribution of the enzymes can have a dramatic effect on the
capacity of push-pull networks to amplify input
signals: the maximum gain of the network in which one enzyme is localized at
one end of the cell, while the other is not, can be much lower than
that of the network in which the components are uniformly distributed
in space. Importantly, this effect occurs over a range of diffusion
constants, protein concentrations, and enzymatic activities that is
typical for living cells.

In the next section, we introduce the push-pull network. In the
Results section, we first present the input-output relations for
both networks. We show that the gain can be much reduced when the
enzymes are spatially separated, and demonstrate that the
magnitude of this effect depends upon the diffusion constants of the
diffusing components. To elucidate the dose-response curves, we discuss in
the subsequent sections the spatial concentration profiles in both the low
 and high activation limits. This analysis reveals that the
maximum gain in the non-uniform system is reduced, because the response of the
network depends on the position in the cell. Interestingly, the
calculations also show that separating the enzymes in space does not only
attenuate strong signals by limiting the maximum response, but can also
enhance the propagation of weak signals. \\[0.3cm]

\noindent {\bf \sf \Large Methods}\\[0.2cm]
\label{Sec2a}
\noindent {\bf \large \sf The push-pull network}\\[0.1cm]
A push-pull network consists of two Michaelis-Menten reactions (see
also Fig.~\ref{fig:fig1}):
\begin{eqnarray}
{\rm E_a}+ {\rm X}\overset{k_1}{\underset{k_2}{\rightleftharpoons}}
{\rm E_aX}\stackrel{k_3}{\rightarrow}{\rm E_a+X^{*}}\\
{\rm E_d}+{\rm
  X}^{*}\overset{k_4}{\underset{k_5}{\rightleftharpoons}}{\rm
  E_dX}^{*}\stackrel{k_6}{\rightarrow}{\rm E_d}+{\rm X}
\end{eqnarray}
Here, ${\rm E_a}$ is the activating enzyme that provides the input
signal, and ${\rm E_d}$ is the deactivating enzyme. The substrate
${\rm X}$ is the unmodified messenger that serves as the detection
component and ${\rm X}^*$ is the modified messenger that provides the
output signal; \EaX\ denotes the activating enzyme bound to its
substrate \X\ and \EdXa\ is the deactivating enzyme bound to its
substrate \Xa. 

If all the components are uniformly distributed in space, then the
chemical rate equations that correspond to this network are: 
\begin{eqnarray}
\frac{\partial [{\rm X}^*]}{\partial t} &=& k_3 [{\rm E_aX}] - k_4 [{\rm E_d}] [{\rm
    X^*}] + k_5 [{\rm EdX^*}]\label{eq:ppu1}\\
\frac{\partial [{\rm X}]}{\partial t} &=& k_6 [{\rm E_dX^*}] - k_1 [{\rm E_a}] [{\rm X}]
 \label{eq:ppu2}  + k_2 [{\rm EaX}] \\
\frac{\partial [{\rm E_a}]}{\partial t} &=& (k_2 + k_3) [{\rm E_aX}] - k_1[{\rm E_a}] [{\rm
    X}]  \label{eq:ppu3}\\
\frac{\partial [{\rm E_aX}]}{\partial t} &=& k_1[{\rm E_a}][{\rm X}] - (k_2 + k_3)
    [{\rm E_aX}]  \label{eq:ppu4}\\
\frac{\partial [{\rm E_d}]}{\partial t} &=& (k_5 + k_6) [{\rm E_dX^*}] - k_4[{\rm E_d}] [{\rm
    X^*}]  \label{eq:ppu5}\\
\frac{\partial [{\rm E_dX^*}]}{\partial t} &=& k_4[{\rm E_d}][{\rm X^*}] - (k_5 + k_6)
    [{\rm E_dX^*}]  \label{eq:ppu6}
\end{eqnarray}
Here, $[...]$ denotes the concentrations of the species. The
steady-state input-output curve of this network can be obtained
analytically~\cite{Goldbeter81}.

We will compare the behavior of this network to that of a network in which
the activating enzyme ${\rm E_a}$ is located at one pole of the cell,
while the other components can freely diffuse in the cytoplasm. The
cell is assumed to be cylindrically symmetric. Since we are interested
in the mean concentration profiles, it is meaningful to integrate out
the lateral dimensions $y$ and $z$. We thus consider a simplified 1D
model, with concentrations as a function of $x$ only. This leads to
the following reaction-diffusion equations:
\begin{eqnarray}
\frac{\partial [{\rm X}^*]}{\partial t} &=& D\frac{\partial^2 [{\rm
    X^*}]}{\partial x^2} + k_3 [{\rm E_aX}] \delta(x)- k_4 [{\rm E_d}] [{\rm
    X^*}]  + \nonumber\\
&& k_5 [{\rm EdX^*}] \label{eq:ppnu1}\\
\frac{\partial [{\rm X}]}{\partial t} &=& D \frac{\partial^2 [{\rm
    X}]}{\partial x^2} + k_6 [{\rm E_dX^*}] - k_1 [{\rm E_a}] [{\rm
    X}] \delta(x)
    \label{eq:ppnu2} + \nonumber \\
&& k_2 [{\rm EaX}] \delta(x)\\
\frac{\partial [{\rm E_a}]}{\partial t} &=& (k_2 + k_3) [{\rm E_aX}] - k_1[{\rm E_a}] [{\rm
    X}](0)  \label{eq:ppnu3}\\
\frac{\partial [{\rm E_aX}]}{\partial t} &=& k_1[{\rm E_a}][{\rm X}](0) - (k_2 + k_3)
    [{\rm E_aX}]  \label{eq:ppnu4}\\
\frac{\partial [{\rm E_d}]}{\partial t} &=& D\frac{\partial^2 [{\rm
    E_d}]} {\partial x^2} + (k_5 + k_6) [{\rm E_dX^*}] - \nonumber \\
&&k_4[{\rm E_d}] [{\rm
    X^*}]  \label{eq:ppnu5}\\
\frac{\partial [{\rm E_dX^*}]}{\partial t} &=& D\frac{\partial^2 [{\rm
    E_dX^*}]}{\partial x^2} + k_4[{\rm E_d}][{\rm X^*}] - \nonumber\\
&& (k_5 + k_6)
    [{\rm E_dX^*}]  \label{eq:ppnu6}
\end{eqnarray}
The components \Ea\ and \EaX\ are localized in the membrane at one end
of the cell; the unit of their concentrations is number of molecules
per area. The other components diffuse in the cell.  Their
concentrations, which are in units of number of molecules per volume,
depend on the position $x$ in the cell, where $x$ measures the
distance from the pole at which \Ea\ and \EaX\ are localized; only in
Eqs.~\ref{eq:ppnu3} and~\ref{eq:ppnu4} is the $x$ dependence
explicitly indicated to emphasize that the \Ea-\X\ association rate
depends on the concentration of \X\ at contact. Zero-flux boundary
conditions are imposed at both cell ends. The steady-state
input-output relations of the network described by
Eqs.~\ref{eq:ppnu1}-\ref{eq:ppnu6} were obtained numerically by
discretizing the system on a (1D) grid and propagating
Eqs.~\ref{eq:ppnu1}-\ref{eq:ppnu6} in space and time until
steady-state was reached.

We consider a cell with the typical dimensions of an {\em E. coli}
cell: the length of the cell, $L$, is thus on the order of $3\mu{\rm
  m}$~\cite{Vaknin04}. We assume the same diffusion constants for all
the components that can diffuse in the cytoplasm. This is for reasons
of simplicity; it is not essential for the main conclusions of our
work. To focus on the effect of enzyme localization on the
input-output relation, we assume for both networks that $k_1 = k_4$,
$k_2 = k_5$, $k_3 = k_6$; the Michaelis-Menten constants for the
modification and demodification reactions are thus the same: $K_{\rm
  M,a} \equiv (k_2 + k_3)/k_1 = K_{\rm M,d} \equiv (k_5 + k_6)/k_4$.
To compare the two networks on equal footing, the total concentration
of activating enzyme, $\cEaT \equiv \cEa + \cEaX$, was chosen such
that ${\rm [E_a]_T^{nu}} = L {\rm [E_a]_T^u}$, where ${\rm
[E_a]_T^{nu}}$ is the concentration (per unit area) in the non-uniform
system and ${\rm [E_a]_T^u}$ is the concentration (per unit volume) in
the spatially uniform network.  This choice ensures that the total number of
activating enzyme molecules in the whole cell is the same for both
systems. In what follows, we will report $\cEaT \equiv {\rm
[E_a]_T^u}$.

In the calculations, we vary the concentration of the activating
enzyme, \Ea, which is the input signal. The total concentration of the
deactivating enzyme, \Ed, is kept constant at $\cEdT = 0.5\mu{\rm M}$;
the rate constants are fixed at $k_1 = k_4 = 10^8\, {\rm
    M}^{-1}{\rm s}^{-1}$, $k_2 = k_5 = 25\, {\rm s}^{-1}$, $k_3 = k_6 = 25
  \,{\rm s}^{-1}$, corresponding to Michaelis-Menten constants of $\KM =
  \KMa = \KMd = 0.5\mu{\rm M}$. We will study extensively the effect
of changing the diffusion constant $D$ and the total substrate
concentration $\cST \equiv \cXT +  \cXaT$, where $\cXT \equiv \cX + \cEaX / L$
is the total concentration of \X\ and $\cXaT \equiv  \cXa + \cEdXa$
is the total concentration of \Xa. Their
base-line parameters, however, are: $D=10\mu{\rm m}^2{\rm s}^{-1}$ and
$\cST = 20\mu{\rm M}$. The magnitude of the diffusion
constant~\cite{Elowitz99}, as well as the values of the
Michaelis-Menten constants, enzyme concentrations, and substrate
concentrations, are typical for
prokaryotic~\cite{Sourjik02_2} and eukaryotic cells~\cite{Ferrell96}.\\[0.3cm]

\noindent {\bf \sf \Large Results}\\[0.2cm]
\noindent {\bf \large \sf The input-output relation}\\[0.1cm]
Goldbeter and Koshland showed that if the antagonistic enzymes in a
push-pull network operate near saturation (see Fig.~\ref{fig:fig1}), a
small change in the concentration of the activating enzyme ${\rm E_a}$
can lead to a large change in the output, the modified messenger ${\rm
X}^*$~\cite{Goldbeter81}. The enzymes become more saturated with
substrate when either the Michaelis-Menten constants $K_{\rm M,a}$ and
$K_{\rm M,d}$ decrease, or the total substrate concentration $[{\rm
S}]_{\rm T} = [{\rm X}] + [{\rm E_a X}]/L + [{\rm E_dX^*}] + [{\rm
X^*}]$ increases.  Fig.~\ref{fig:fig2}a shows the steady-state
input-output relation for a push-pull network in which all the
components are uniformly distributed in space, for different substrate
concentrations. It is seen that as the substrate concentration is
increased, the sharpness of the response is drastically enhanced. This
is the hallmark of the mechanism of zero-order ultrasensitivity.

In many systems, such as the bacterial chemotaxis network
of {\em E. coli}~\cite{Vaknin04}, the two antagonistic enzymes are
colocalized at the same pole, while the detection component ${\rm X}$
and the messenger ${\rm X^*}$ can diffuse through the
cytoplasm. While the time-dependent response curves of such a network
will differ from those of the two networks considered here, the steady-state
dose-response curves will be identical to those of a network in which all
the components are homogeneously distributed in the cytoplasm. The
response curves shown in Fig.~\ref{fig:fig2}a thus also pertain to
push-pull networks in which the two enzymes are colocalized at one end
of the cell, while their substrates freely diffuse in the
cytoplasm. Indeed, also in these networks the mechanism of zero-order
ultrasensitivity can strongly amplify input signals.

{\bf \sf Spatially separating the enzymes reduces the gain.}
Fig~\ref{fig:fig2}b shows the dose-response curves for a push-pull
network in which the activating enzyme ${\rm E_a}$ is localized at one
pole of the cell, while the other components diffuse in the
cytoplasm. Three points are worthy of note. The first is that the
maximum output signal, the concentration of the messenger ${\rm X^*}$,
is much lower than that of the corresponding network in which all
components are uniformly distributed in space (see
Fig.~\ref{fig:fig2}). In fact, while in the spatially uniform network,
the fraction of modified substrate, $\cXaT /\cST$, always approaches
unity if ${[\rm E_{a}]_T/[E_d]_T}$ becomes large, in the non-uniform
network the fraction of modified substrate saturates to a lower level:
even when the concentration of activating enzyme is much higher than
that of the deactivating enzyme, not all substrate ${\rm X}$ is
converted into ${\rm X}^*$.  The second point to note is that as the
total substrate concentration decreases, the inflection point of the
dose-response curve shifts to lower values of ${\rm
  [E_a]_T/[E_d]_T}$. The last, and perhaps most important, point to
note is that the {\em sharpness} of the response of the network is
much weaker than that of the network in which the enzymes are either
colocalized or uniformly distributed in space. The insets of Fig.
\ref{fig:fig2} shows the logarithmic gain, $g\equiv\partial
  \ln\cXaT/\partial \ln \cEaT$, as a function of $\cEaT/\cEdT$ for both
  networks ($\cEdT$ is kept constant). It is seen that for both low $\cEaT/\cEdT$ and high
  $\cEaT/\cEdT$ the gain is small and fairly similar for both
  networks, while for the symmetric networks considered here, at
  $\cEaT \approx \cEdT$ the gain is maximal, but smaller for the
  network in which the enzymes are spatially separated. Hence, spatially
  separating the two antagonistic enzymes reduces the
  maximum gain of a push-pull network.

{\bf \sf The dose-response curves strongly depend on the diffusion constants.}
The extent to which the spatial separation of the opposing enzymes
can change the response of the network depends on the magnitude of the
diffusion constant of the components. This is illustrated in
Fig.~\ref{fig:fig3}. This figure shows the input-output relation of a
push-pull network where the activating enzyme is located at one end of
the cell, while the other components diffuse freely in the cytoplasm,
for different values of the diffusion constant. This network is in the
zero-order regime: the total substrate concentration is large
compared to the concentrations of the enzymes and the Michaelis-Menten
constants. Yet, for low values of the diffusion constants, the
response is rather weak. As the diffusion constant increases, however,
the sharpness of the response markedly increases. For $D\rightarrow
\infty$, the input-output relation approaches that of a push-pull
network in which all components are either uniformly distributed in space, or
colocalized in one region of the cell.

{\bf \sf Spatially separating the enzymes attenuates the propagation
  of strong signals, but can enhance the transmission of weak
  signals.}  Fig.~\ref{fig:fig3} shows that in a zero-order network in
which only the activating enzyme is localized at one pole of the cell,
the concentration of \Xa\ decreases with decreasing diffusion constant
when $\cEaT > \cEdT$, but {\em increases} with decreasing diffusion
constant when $\cEaT < \cEdT$. This means that when the input signal
is strong (high kinase activity), the response of a network in
which the enzymes are spatially separated is weaker than that of a
network in which the enzymes are either uniformly distributed or
colocalized in space; conversely, when the input signal is weak (low
kinase activity), the
spatially non-uniform network can respond more strongly than a uniform
network.  Spatially separating the antagonistic enzymes will
thus attenuate strong input signals, but can also amplify weak input signals.\\[0.2cm]

\noindent {\bf \sf \large Mechanism: concentration gradients}\\[0.1cm]
To explain the effect of enzyme localization on the amplification
properties of push-pull networks, it is instructive to
consider the effect of diffusion on the input-output relation: in the
limit that $D\rightarrow\infty$, the response of the network in which
the activating enzyme is located at the pole, while the other is
distributed in the cytoplasm, approaches that of a network in which
the enzymes are either uniformly distributed in space or colocalized
at the pole. The effect of diffusion on the response
curves can be understood by considering the effects of diffusion on the
spatial concentration profiles.

In a push-pull network where the antagonistic enzymes are either
uniformly distributed or colocalized in space, the
steady-state spatial concentration profiles of the freely diffusing
components are uniform across the cell. In a push-pull network where
the two antagonistic enzymes are spatially separated, concentration
gradients of the freely diffusing components can
form. Fig.~\ref{fig:fig4} shows for a zero-order network in which the activating
enzyme is located at one pole of the cell, while the other is not, the
concentration profiles of \Xa\ and \EdXa, for
three different (total) concentrations of the activating enzyme \Ea,
$\cEaT$. The concentrations of \Xa\ and \EdXa\ are highest near the
pole where \X\ is activated, and decay in the cytoplasm where \Xa\ is
deactivated. Moreover, the concentration profiles increase as $\cEaT$
increases. These gradients impose fundamental
limits on the maximum gain of the system.

To clarify the effect of diffusion on the concentration profiles
and the input-output relations, it is instructive to recall that,
in general, the spatio-temporal evolution of $\cXa$ is given by the
interplay of activation, deactivation and diffusion of \Xa:
\begin{equation}
\label{eq:dXadt}
\frac{\partial \cXa}{\partial t} = D\frac{\partial^2 [{\rm
    X^*}]}{\partial x^2} + J \delta (x) - \gamma (x).
\end{equation}
Here, $J$ denotes the influx of \Xa\ into the system, while $\gamma$
denotes the deactivation rate of \Xa\ at position $x$. If the
formation of the enzyme-substrate complexes is fast (see {\em
  Text S1}), then $J$ and $\gamma$ are given by
\begin{eqnarray}
J &=& k_3 \cEaT L \frac{\cX(0)
}{\KMa+\cX(0)} \label{eq:J}\\
 \gamma(x)& =& k_6 \cEdT
  \frac{\cXa (x)}{\KMd+\cXa (x)} \label{eq:gamma}.
\end{eqnarray}
Here, $\cEaT \equiv \cEa + \cEaX$ and $\cEdT \equiv \cEd + \cEdXa$ are
the total concentrations of \Ea\ and \Ed, respectively. Combining
Eq.~\ref{eq:ppnu5} with Eq.~\ref{eq:ppnu6} reveals that the total
concentration profile of \Ed, $\cEdT (x)$, is constant in space if, as
assumed here, the diffusion constants of the enzyme \Ed, and that of
the enzyme bound to its substrate, \EdXa, are the same.  The synthesis
rate of \Xa\ depends upon the concentration of \X\ at contact and hence upon the
concentration of \Xa\ at contact; similarly, the deactivation rate of
\Xa\ at position $x$ depends upon the concentration of \Xa\ at
$x$. This is important to note, because, as we discuss below, the
dose-response curves are determined by the sensitivities of the influx
$J$ and the deactivation rate $\gamma$ to changes in the substrate
concentration.  We will now first discuss the input-output
relations of zero-order push-pull networks, and then briefly the
response curves of push-pull networks
that are in the linear regime.\\

\noindent{\bf \large \sf Push-pull networks in the zero-order regime}\\
Figs.~\ref{fig:fig4}-\ref{fig:fig6} show the concentration profiles
and dose-response curves of push-pull networks
that are in the zero-order regime. We now discuss the limits of weak
and
strong activation separately.\\[0.1cm]

\noindent {\bf \sf Weak activation}\\
We first consider the regime in which the concentration of the
activating enzyme is lower than that of the deactivating enzyme,
corresponding to Figs.~\ref{fig:fig5}a-c.  In the limit that $\cEaT
\ll \cEdT$, $\cX$ will be large and $\cXa$ will be small. As a
consequence, \Ea\ is saturated with its substrate \X, while \Ed\ is
not saturated with its substrate \Xa. Because \Ea\ is saturated, the influx
of \Xa\ into the system is constant (i.e. independent of $\cX$ and $\cXa$)  and given by $J = k_3 \cEaT L$ (see
Eq.~\ref{eq:J}). Because \Ed\ is unsaturated, the deactivation
rate $\gamma$ is proportional to $\cXa$: $\gamma(x) = \mu \cXa(x)$,
with $\mu = k_6 / \KMd \cEdT$ (see Eq.~\ref{eq:gamma}). This means
that in this regime the deactivation rate per particle is constant.

With the influx $J$ being constant and the deactivation rate $\gamma$
being proportional to $\cXa$, Eq.~\ref{eq:dXadt} can be solved
analytically (see {\em Text S1}). 
Defining the characteristic decay length of
\Xa\ to be $\lambda = \sqrt{D/\mu}$, then, if $L \gg \lambda$, the
solution is
\begin{equation}
\label{eq:cXa_xEalow}
\cXa(x) = \frac{J\lambda}{D}\exp(-x/\lambda).
\end{equation}

Eq.~\ref{eq:cXa_xEalow} reveals that when $D$ increases, the profile
decays more slowly, and the concentration of \Xa\ at contact decreases. When $D$
increases, the \Xa\ molecules diffuse away from the pole more
rapidly. Because in the regime considered here, namely $\cEaT \ll
\cEdT$, the influx of \Xa\ is constant and independent of $D$, the
concentration of \Xa\ close to the pole will decrease when the molecules
diffuse away more rapidly, while the
concentration further away will increase.  In fact, in this limit the total concentration of \Xa, $\cXa^{\rm cell}$, is independent of
the diffusion constant; this can be verified by integrating
Eq.~\ref{eq:cXa_xEalow} over the whole cell, which yields $\cXa^{\rm cell} =
J/\mu$. The fact that the total concentration of $\cXa$ is independent of
the diffusion constant, means that the response of the network does
not depend upon the spatial distribution of the enzymes.


When $\cEaT$ increases, $\cXa$ increases and $\cX$ decreases. As a
result, \Ea\ becomes less saturated, while \Ed\ becomes
saturated. Hence, the influx $J$ will at some point become sensitive
to \X, while the deactivation rate $\gamma$ will no longer be
proportional to $\cXa$. However, in the zero-order regime considered
here, the total substrate concentration $\cST$ is large as compared to
the enzyme concentrations and the Michaelis-Menten constants. This
means that as $\cEaT$ is raised, such that $\cX(0)$ decreases and
$\cXa(0)$ increases, initially \Ea\ will remain fully saturated, while \Ed\
will become saturated.  This implies that there is a
range of \Ea\ concentrations, where the influx $J$ is still constant,
but the deactivation rate $\gamma$ is no longer proportional to
$\cXa$. In this regime, the concentration of \Xa\ increases with
decreasing diffusion constant. Indeed, in this range, where $\cEaT <
\cEdT$, the spatially non-uniform network will respond stronger than
the spatially uniform network (see Figs.~\ref{fig:fig3}
and~\ref{fig:fig5}a-c).

The significance of the diffusion term for $\cEdXa$ in
Eq.~\ref{eq:ppnu6} impedes a transparent analytical derivation of
$\cXa(x)$ in this regime (see {\em Text S1}). However, under
the condition that the influx $J$ is constant, we can prove that the
total amount of \Xa\ must decrease with increasing diffusion constant
when $\gamma$ is no longer proportional to $\cXa$. The proof can be
found in {\em Text S1}. 

Here we give a more intuitive explanation for the observation that a
network in which the enzymes are spatially separated can respond
stronger than a network in which the enzymes are similarly distributed
in space.  Ultimately, it is a consequence of the non-linear
enzyme-substrate binding curve and the resulting hyperbolic dependence
of the deactivation rate $\gamma$ on $\cXa$ (see
Eq.~\ref{eq:gamma}). More specifically, this effect can arise when
  the diffusion constant is low and/or the deactivating enzyme
   operates close to, but not at, saturation in the uniform system; in
  this uniform system, \Xa\ is distributed evenly through the
  cell and all particles \Xa\ experience the same deactivation
  rate $\mu$. In the spatially non-uniform system, $\cXa$ is
  higher near the pole. If all the deactivating enzyme molecules would
  operate in the linear regime, i.e. if all deactivating enzyme
  molecules would not be saturated, then all particles \Xa\ would
  still experience the same degradation rate $\mu$; in this scenario,
  the increase in the number of \Xa\ particles close to the pole
  would precisely balance the decrease in the number of \Xa\ 
  particles further away from the pole, as compared to the uniform
  network. However, if the concentration of the deactivating enzyme
  with respect to that of its substrate is lower, i.e. if the enzyme
  operates close to saturation, then the scenario can arise that the
  deactivating enzyme molecules near the pole become saturated (Fig.
\ref{fig:fig5} b), while in the corresponding uniform network they
  are not.  In this scenario, the \Xa\ particles that are located
  close to the pole in the non-uniform network experience a lower
  effective deactivation rate than the \Xa\ particles in the
  spatially uniform network. This will enhance the response of the
  non-uniform system as compared
  to that of the uniform system.\\[0.1cm]

\noindent {\bf \sf Strong activation}\\
We now discuss the effect of the diffusion speed on the concentration
profiles of \Xa\ when $\cEaT > \cEdT$ (see Figs.~\ref{fig:fig5}d-f).
In this regime, $\cX$ is low and $\cXa$ is high. This reverses the
saturation behavior of the antagonistic enzymes: while in the
weak-activation limit \Ea\ is saturated and \Ed\ is unsaturated, now
\Ea\ is unsaturated and \Ed\ is fully saturated. This also reverses the
sensitivities of the influx $J$ and the deactivation rate $\gamma$ to
changes in the substrate concentration. Indeed, in the
strong-activation regime not the
influx $J$ is constant, but
rather the deactivation rate: $\gamma = k_6 \cEdT$ (see
Eq.~\ref{eq:gamma}). 

In the limit that the deactivation rate $\gamma$ is constant,
Eq.~\ref{eq:dXadt} can be solved in steady state (see {\em
  Text S1}). The solution is
\begin{equation}
\label{eq:cXa_x_Eahigh}
\cXa(x) = c_0 + c_1 x + \frac{1}{2} c_2 x^2,
\end{equation}
where $c_2 = k_6 \cEdT / D$, $c_1 = -k_6 \cEdT L / D$, and $c_0 =
\cXa(0) = \cST - \cEdT(1+k_6/k_3(1+\KMa/\cEaT))$.  It is seen that in
the high-activation regime, the concentration profile decays
algebraically, rather than exponentially, as in the limit of weak
activation. This is precisely because in the high-activation regime
the total deactivation rate $\gamma$ is constant in space, while in
the weak-activation limit $\gamma$ is proportional to the concentration of
\Xa, which varies in space. In fact, in the weak-activation
limit the deactivation rate {\em per particle} is constant in space
and equal to $\mu$. In contrast, in the strong-activation regime, the
deactivation rate per particle is not only lower than $\mu$ on
average, but also varies in space: the higher $\cXa$ as compared to
$\cEdT$ (which is constant in space and sets the total deactivation
rate), the lower the deactivation rate per particle; activated
particles close to the pole thus experience a
lower deactivation rate and hence travel further on average before
they are deactivated.

The expression for $c_0 = \cXa(0)$ reveals that as $\cEaT$ increases,
the concentration of \Xa\ close to the pole where \Ea\ is located,
increases. In the limit that $\cEa \rightarrow \infty$, $\cXa(0)
\rightarrow \cST - 2 \cEdT \approx \cST$, which means that {\em close}
to the pole of the cell where the activating enzyme is located, all
the substrate \X\ is converted into \Xa\ and \EdXa\ (see Fig.~\ref{fig:fig5}d).
Importantly, Fig.~\ref{fig:fig5}d also shows that as the distance from
the pole increases, the fraction $\cXaT(x) / \cST(x)$ decreases, {\em
even in the limit that} $\cEaT \gg \cEdT$. When $\cEaT \gg \cEdT$, all
the substrate molecules at the pole will indeed be modified. However,
these molecules will then diffuse away from the pole into the
cytoplasm, where they can be demodified by the deactivating enzyme
molecules, but not remodified. Hence, when the activating enzyme is
spatially separated from the deactivating enzyme, it will never be
possible to convert all the substrate molecules in the system (see
Fig.~\ref{fig:fig2}). This is in marked contrast with the situation in
which the activating and deactivating enzymes are not spatially
separated. In this case,
all substrate molecules can be converted into \Xa\ when $\cEaT \gg
\cEdT$ (see Fig.~\ref{fig:fig2}).

The expression for $c_0 = \cXa(0)$ also reveals that in the limit that
$\cEaT \gg \cEdT$, the concentration of \Xa\ at $x=0$ is independent of
the diffusion constant.  However, while $\cXa (0)$ does not depend on the diffusion constant, the rate at which
$\cXa(x)$ decays with the distance from the pole, does depend on it. Eq.~\ref{eq:cXa_x_Eahigh}, with $c_1 = -k_6\cEdT L/D$,
shows that the concentration profile of \Xa\ decays more slowly when
the diffusion constant increases (see also
Fig.~\ref{fig:fig5}d). These two observations, when taken together,
imply that the total concentration of \Xa\ in the whole system
increases with increasing diffusion constant. This can be verified by
integrating Eq.~\ref{eq:cXa_x_Eahigh} over the length of the cell,
which gives $\cXa^{\rm cell} \sim a
- b/D$, where $a$ and $b$ are positive constants. 

These results can be understood by comparing the influx of \Xa\ with
the efflux of \Xa. When $\cEaT > \cEdT$, the deactivation rate is
constant and hence independent of the diffusion constant. Since the
total deactivation rate of \Xa\ is independent of the diffusion
constant, the total influx of \Xa, which in steady state must balance
the total efflux by deactivation, is also independent of the diffusion
constant. The influx of \Xa\ depends on $\cEaX$ and thus on the
concentration of \Xa\ at $x=0$, as discussed above. Hence, the
concentration of \Xa\ at $x=0$ must be independent of the diffusion
constant. A more intuitive explanation is as follows: As the diffusion
constant increases, the \Xa\ molecules will diffuse away from the pole
more rapidly. This would tend to lower the concentration of \Xa\ at
$x=0$. However, this process is accompanied by an increase in the flux
of \X\ towards the pole ($\cST(x)$ is constant); because in the strong
activation limit \Ea\ is unsaturated, this would tend to increase
$\cEaX$ and thereby the influx of \Xa, which would raise the
concentration of \Xa. In steady state, these processes balance each
other such that the concentration of \Xa\ at contact does not depend
on the diffusion constant. However, while $\cXa(0)$ does not change
with the diffusion constant, the \Xa\ molecules do diffuse away from
the pole more rapidly when the diffusion constant increases. This
means that the total concentration profile of \Xa\ must increase with
increasing diffusion constant (see Fig.~\ref{fig:fig3}). Indeed, only
in the limit that $D\rightarrow \infty$ and $\cEaT \gg \cEdT$, can all
the substrate molecules be converted into \Xa\ (see
Fig.~\ref{fig:fig3} and Fig.~\ref{fig:fig5}d). In the
strong-activation limit, spatially separating the antagonistic enzymes
thus always weakens the response,
in contrast to the behavior in the weak-activation limit.

It should also be noted that the decay length of the
concentration profile of \Xa, given by $c_1 = -k_6 \cEaT L/D$, does
not only depend upon the diffusion constant, but also upon the
activity of the deactivating enzyme. In the spatially non-uniform
system, the maximum response (i.e. the response when $\cEaT \gg
\cEdT$) decreases as the catalytic activity of the deactivating enzyme,
$k_6$, increases. The reason is that the \Xa\ molecules will
travel a shorter distance before they are deactivated, when the
deactivation rate is higher. The extent to which spatially separating
the enzymes weakens the maximum response thus depends upon both the
diffusion constant and the deactivation rate of the \Xa\ molecules.\\

\noindent{\bf \sf Space-dependent amplification}\\[0.1cm]
Fig.~\ref{fig:fig6} shows that if
the activating enzyme is localized at one pole of the cell, while the
deactivating enzyme can freely diffuse through the cytoplasm,  the
response of the network will depend upon the position in the
cell.  As can be deduced from Figs.~\ref{fig:fig4}
and~\ref{fig:fig5}, $\cEdXa$ depends significantly on the position in
the cell when $\cEaT < \cEdT$. When $\cEaT> \cEdT$, however, $\cEdXa$
becomes virtually independent of the position $x$, because then all
the deactivating enzyme molecules are saturated. The opposite trend is
observed for $\cXa$: when $\cEaT < \cEdT$, $\cXa$ is low everywhere in
the cell, while if $\cEaT > \cEdT$, $\cXa$ strongly depends upon the
position in the cell. The reason is, as discussed in the previous
section, that even when $\cEaT \gg \cEdT$, not all \X\ can be
converted into \Xa\ if the two antagonistic enzymes are spatially
separated.

Interestingly, the average response of $\cEdXa$ in the spatially
non-uniform system is very similar to that in the system in
which the two enzymes are not spatially separated. Yet, the response
of $\cXa$ does differ markedly between the two systems. This is a
result of the strong non-linearity in the amplification mechanism of
zero-order ultrasensitivity: because the activation and deactivation
reactions are zero-order in the substrate concentrations $\cX$ and
$\cXa$, respectively, even when $k_3 \cEaX$ is only marginally larger
than $k_6 \cEdXa$, predominantly all \X\ molecules will be converted
into \Xa\ \cite{Ferrell96}.

Lastly, Fig.~\ref{fig:fig6}a shows that the inflection point of the
dose-response curve  depends on the position $x$ in the
cell. The inflection point shifts to higher $\cEaT/\cEdT$ as the
distance from the anterior pole increases; this effect becomes more
pronounced as $D$ decreases (data not shown). The fact that the
inflection point depends on the position $x$ is one of the principal
reasons why the response in
the spatially non-uniform system is weaker than that of the uniform system.\\[0.2cm]

\noindent{\bf \sf \large Push-pull networks in the linear regime}\\
Push-pull networks in living cells are not always in the zero-order
regime~\cite{Ferrell96,Sourjik02_2}. In the linear regime, push-pull
networks do not amplify signals, but can enhance the reliability of
cell signaling by making it robust against fluctuations in the
concentrations of the components due to noise in gene expression~\cite{Kollmann05}. It
is therefore meaningful to study how the input-output relation of a
push-pull network in the linear regime depends upon the spatial
distribution of the antagonistic enzymes. A push-pull network in the
linear regime is given by:
\begin{eqnarray}
{\rm E_a} + {\rm X}  \stackrel{k_1}{\rightarrow} {\rm E_a} + {\rm {X}^*}\\
{\rm E_d} + {\rm X}^* \stackrel{k_2}{\rightarrow} {\rm E_d} + {\rm X}
\end{eqnarray}
The steady-state concentration profiles for these linear push-pull
networks can be derived analytically.

The principal result is that for push-pull networks that are in the
linear regime, spatially separating the antagonistic enzymes {\em
  always weakens the response}. This can be seen by comparing the
response curve for $\cST = 0.4 \KM$ in Fig.~\ref{fig:fig2}a with that
in Fig.~\ref{fig:fig2}b. The reason why for linear networks spatially
separating the enzymes reduces the response in the strong-activation
limit is the same as that for zero-order networks. The reason that, in
contrast to zero-order networks, also in the weak-activation limit the
response is weakened, is more subtle. In zero-order networks that are
in the weak-activation limit, \Ea\ is saturated and, consequently, the
influx $J$ is independent of the concentration of \X\ at the pole. In
linear networks, \Ea\ is unsaturated and the influx $J$ is
proportional to $\cX(0)$. As $D$ decreases, $\cXa(0)$ tends to
increase and $\cX(0)$ tends to decrease ($\cST(x)$ is constant
in space). Because in the linear regime $J$ is proportional to
$\cX(0)$, this would lower the influx of \Xa, which, in turn, would lower the
concentration of \Xa. Spatially separating the antagonistic enzymes
thus amplifies weak signals if the push-pull
network operates in the zero-order regime, but not in the linear regime.\\[0.3cm]

\noindent {\bf \sf \Large Discussion}\\[0.2cm]
{In a push-pull network that operates deeply in the zero-order
  regime, the activation rate is given by $k_3\cEaT$, while the
  deactivation rate is given by $k_6 \cEdT$; both rates are thus
  independent of the substrate concentration. If both enzymes are
  uniformly distributed, or colocalized, then essentially all
  substrate molecules will be activated when $k_3 \cEaT > k_6 \cEdT$,
  while they will be predominantly deactivated when $k_3 \cEaT < k_6
  \cEdT$. To drive the modification reactions to completion, it is
  indeed essential that the antagonistic enzymes are not spatially
  separated. If the antagonistic enzymes are separated, then the
  enzyme with the lower global activity, can locally still have a
  higher activity than the other enzyme. More in general, spatially
  separating the enzymes means that the balance between activation and
  deactivation depends upon the position in the cell, and this
  ``smearing'' of the response always tends to reduce the sharpness of
  the global response curve.  

If information about changes in the environment has to be
  transmitted, then the gain---the change in the output divided by the
  change in the input---is a critical quantity. In fact, the {\em
  maximum} gain is then usually the most relevant quantity,
  because signaling networks are often tuned to this point of maximum
  gain: the input-output function of a module and the concentration of
  its input signal are often optimized with respect to each other. The
  intracellular chemotaxis network of {\em E. coli} provides
  a clear example: the steady-state intracellular concentration
  of the messenger CheYp is around $3 \mu{\rm M}$, which is precisely
  the concentration at which the flagellar motors respond most
  strongly. Our analysis shows that from the perspective of signal
  amplification, the best strategy is to either colocalize the
  antagonistic enzymes or to uniformly distribute them in space:
  spatially separating the enzymes always weakens the maximum
  response.

Nevertheless, as mentioned in the introduction, spatial gradients
  of messenger proteins are often observed. Indeed, maximizing the
  gain is not the only design principle in cell signaling. Firstly,
  while in some cases, such as {\em E. coli} chemotaxis, the
  signal has to be transmitted to a large number of places throughout
  the cell's cytoplasm or membrane \cite{Rao05}, in other cases
  the signal has to be transmitted to distinct regions, such as the
  nucleus, or be confined to a small region near the membrane, as in
  the yeast pheromone response where the shmoo tip has to be formed
  locally; in this scenario, spatial gradients might be important,
  since they allow the cell to confine signaling to a narrow domain
  below the cell membrane \cite{Brown99_2,Kholodenko06}. Secondly, a
  sharp response may not always be desirable. In order to respond
  strongly to changes in the input signal over a broad range of input
  signal strengths, the cell does not only need a sharp response
  curve, but it also needs to develop elaborate adaptation mechanisms that can
  reset the network to the point of maximum gain. In {\em
  E. coli}, for instance, the methylation and demethylation
  enzymes CheR and CheB continually adjust the activity of the
  receptor cluster, such that the steady-state intracellular CheYp
  concentration is at $3\mu{\rm M}$. A weaker response curve, however,
  would allow the cell to have a reasonable working range without
  adaptation mechanisms. In this scenario not only the maximum gain
  would be important, but, in fact, the full response curve. Thirdly,
  it might not always be possible to maximize signal amplification by
  optimizing the input-output function of a module with respect to its
  incoming signal, because, for instance, the downstream module also
  has to respond to other incoming signals, while the signal also has
  to act on other downstream modules; the yeast MAPK pathways, which
  exhibit cross talk, provide a prominent example of such a
  scenario. It seems likely that in this case the full response curve,
  with the absolute concentrations of the components, is important. In
  this context it is interesting to note that spatially separating the
  antagonistic enzymes weakens strong signals by reducing the maximum
  output signal (Figs. \ref{fig:fig3} and \ref{fig:fig5}a),
  while it can enhance weak signals if the network operates in the
  zero-order regime (Figs.}  \ref{fig:fig3} and
\ref{fig:fig5}d). This dependence of the input-output relation on
  the spatial distribution of the antagonistic enzymes could be
  exploited by cells to relay different environmental signals
  specifically.

The analysis performed here is essentially a mean-field analysis. It
is assumed that the concentrations are large and that fluctuations can
be neglected. However, in the living cell, the concentrations are often
low, which means that fluctuations can be important. This is
particularly relevant for push-pull networks. Their high gain not only
amplifies the mean of the input signal, but will also amplify the
noise in the input signal~\cite{Paulsson04,Shibata05,TanaseNicola06}. Moreover, when
the modification reactions become more zero order, the intrinsic
fluctuations of the push-pull network, i.e. noise resulting from the
modification reactions themselves, will also
increase~\cite{Berg00}. In fact, it has been shown that when
push-pull networks operate deeply in the zero-order regime,
fluctuations can lead to a bimodal response~\cite{Samoilov05}. All
these analyses of the effect of noise on the amplification mechanism
of zero-order ultrasensitivity have been performed under the
assumption that the enzymes are uniformly distributed in the
cytoplasm. It would clearly be of interest to study the effect of enzyme
(co)-localization on the noise characteristics of push-pull
networks.

Finally, could our predictions be tested experimentally?  To test
our predictions, one would ideally like to perform an experiment on a
system with a canonical push-pull network in which all the
parameters---concentrations of components, rate constants, diffusion
constants---are kept constant, except for the spatial location of one
of the enzymes. This clearly seems a very difficult experiment to
perform, and to our knowledge, no such experiment has been performed
yet, with the possible exception of the experiment by Vaknin and Berg
\cite{Vaknin04}. Vaknin and Berg studied the effect of
phosphatase localization on the response of the intracellular
chemotaxis network of {\em E. coli} cells. This network has a
topology that is very similar to that of the canonical push-pull
networks considered here, and it is believed that in the
wild-type cells both the kinase and the phosphatase are localized at
the cell pole. Vaknin and Berg compared the response of wild-type
cells to that of mutant cells, in which the phosphatase was mutated
such that it freely diffuses in the cytoplasm. They found that the
spatial distribution of the phosphatase can have a marked effect on
the the sharpness of the response, which seems to
support the principal conclusion of our analysis. We would like to
emphasize, however, that to assess the importance of the spatial
distribution of the antagonistic enzymes in a push-pull network, a
careful, quantitative analysis of the network is required. First of
all, our analysis shows that both the quantitative and qualitative
consequences of enzyme localization, depend upon the regime in which
the network operates. For instance, our calculations reveal that if
the activation rate is independent of the messenger concentration, and
if the deactivation rate is linear in the messenger concentration,
then the localization of the phosphatase should have no effect at all
on the response curve. Secondly, it is quite possible that in the
mutant cells not only the spatial distribution of the enzymes is
different, but also their expression level, or even other parameters
such as rate constants. In fact, experiments by Wang and Matsumura
suggest that the activity of the phosphatase in the {\em E. coli}
chemotaxis network is enhanced at the receptor cluster
\cite{Wang96}. Clearly, different rate constants would also tend
to change the response curve of the mutant cells with respect to that
of the wild-type cells. To elucidate the effect of enzyme localization
on the dose-response curve of a network, thus requires quantitative
experiments and quantitative modeling. In a future publication, we
will present a detailed analysis on the importance of phosphatase
localization in the chemotaxis network of{\em E. coli}.

We thank Howard Berg, Dennis Bray, Viktor Sourjik and Ady Vaknin for useful
discussions and Rhoda Hawkins, Ady Vaknin and Viktor Sourjik for a critical
reading of the manuscript. This work is part of the research
program of the ``Stichting voor Fundamenteel Onderzoek der Materie
(FOM)", which is financially supported by the ``Nederlandse
organisatie voor Wetenschappelijk Onderzoek (NWO)''.

\newpage
\begin{center}
\textbf{LIST OF FIGURES}
\end{center}

\begin{enumerate}
\item
A push-pull network. Two enzymes, ${\rm E_a}$ and ${\rm
  E_d}$, covalently (de)modify the components ${\rm X}$ and ${\rm X}^*$,
  respectively. The activating enzyme ${\rm E_a}$ provides the input
  signal, the unmodified component ${\rm X}$ is the detection
  component and the modified component ${\rm X}^*$ provides the output
  signal.
\label{fig:fig1}

\item The input-output relation of the push-pull network shown in
  Fig.~\ref{fig:fig1} as a function of the total substrate
  concentration $[{\rm S]_T}$, for the case in which all components
  are uniformly distributed in space (a) and for the case in which the
  activating enzyme is located at one end of the cell, while the other
  components can diffuse freely through the cell (b). Here, $\cXaT/\cST =
  \int_0^L dx \cXaT (x)/\int_0^L dx \cST (x)$. In (a) and (b),
  ${\rm [E_d]_T} = 0.5 \mu{\rm M}$,  $K_{\rm M,a} = K_{\rm M,d} =
  0.5 \mu{\rm M}$, and $k_3 = k_6 = 25 {\rm s}^{-1}$. In (b), the diffusion constant is $D=10\mu{\rm
    m}^2\rm{s}^{-1}$. The inset shows the logarithmic gain $g
    \equiv \partial \ln\cXaT/ \partial \ln\cEaT$. It is seen that the
  sharpness of the response increases markedly with increasing
  substrate concentration when all the components are uniformly
  distributed in space (a), but much less so when the activating enzyme ${\rm
    E_a}$ is located at one pole of the cell, while the deactivating
  enzyme ${\rm E_d}$ is distributed in the cytoplasm. When both
  enzymes ${\rm E_a}$ and ${\rm E_d}$ are located at one pole, the
  steady-state dose-response curve is identical to that in
  (a).\label{fig:fig2}

\item
The input-output relation of a network in which
  the activating enzyme is located at one pole, while the other
  components can freely diffuse in the cytoplasm, for different values
  of the diffusion constant $D$ (in $\mu{\rm m}^2{\rm s}^{-1}$) of the cytoplasmic components.  The
  inset shows the logarithmic gain $g \equiv \partial
  \ln\cXaT/ \partial \ln\cEaT$. It is seen that
  the gain of the push-pull network strongly increases with increasing
  diffusion constant. If $D \rightarrow \infty$, the dose-response curve
  approaches that of the push-pull network in which the components are
  uniformly distributed in space (and that of the network in which
  the enzymes are colocalized). The total
  substrate concentration is ${\rm [S]_T} = 20 \mu{\rm M}$, the total
  concentration of the deactivating enzyme is ${\rm [E_d]_T} =
  0.5 \mu{\rm M}$, the Michaelis-Menten constants are $K_{\rm M,a} =
  K_{\rm M,d} = 0.5 \mu{\rm M}$, and the catalytic rate constants are $k_3 = k_6 = 25 {\rm s}^{-1}$. \label{fig:fig3}

\item
The concentration profiles for \Xa\ (a) and \EdXa\ (b) in a
  push-pull network in which the activating enzyme is located at one
  pole of the cell, while the other components are distributed in the cytoplasm,
  for three different concentrations of the activating enzyme.  For
  all curves, $\cST = 20\mu{\rm M}$, $\cEdT = 0.5\mu{\rm M}$, $\KMa =
  \KMd = 0.5 \mu{\rm M}$, $k_3 = k_6 = 25 {\rm s}^{-1}$, and $D= 10 \mu{\rm m}^2{\rm
  s}^{-1}$.
\label{fig:fig4}

\item
Profiles of ${\rm [X^*]}$ (a and d),
  ${\rm [E_dX^*]}$ (b and e) and ${\rm [E_d]}$ (c and f). a - c:
  Low concentration of activating enzyme, $\cEaT = 0.5 \cEdT$; d-f: High
activating enzyme concentration, $\cEaT = 1.5 \cEdT$. For the other parameter values, see
  Fig.~\ref{fig:fig4}. \label{fig:fig5}

\item Dose-response curves of the push-pull network in which the
  activating enzyme is localized at one pole of the cell, while the
  other components diffuse in the cytoplasm, for different positions
  in the cell ($x=0$ corresponds to the black left most curve, while
  $x=3\mu{\rm m}$ corresponds to the black right most curve); a) profiles
  of $\cXa$ and b) profiles of $\cEdXa$; note that the response
  becomes sharper further away from the pole. The green curves
  correspond to the average or integrated response of the non-uniform
  system, while the red curves correspond to the uniform
  system. The inset shows the logarithmic gain $g\equiv \partial
    \ln \cXa /\partial \ln\cEaT$ at the respective positions in the
    cell ($x=0,1,2,3, \mu{\rm m}$).  For the parameter values, see
  Fig.~\ref{fig:fig4}.\label{fig:fig6}

\end{enumerate}

\newpage
\vspace*{2cm}
\begin{figure}[h]
\centering \includegraphics[width=0.75\linewidth]{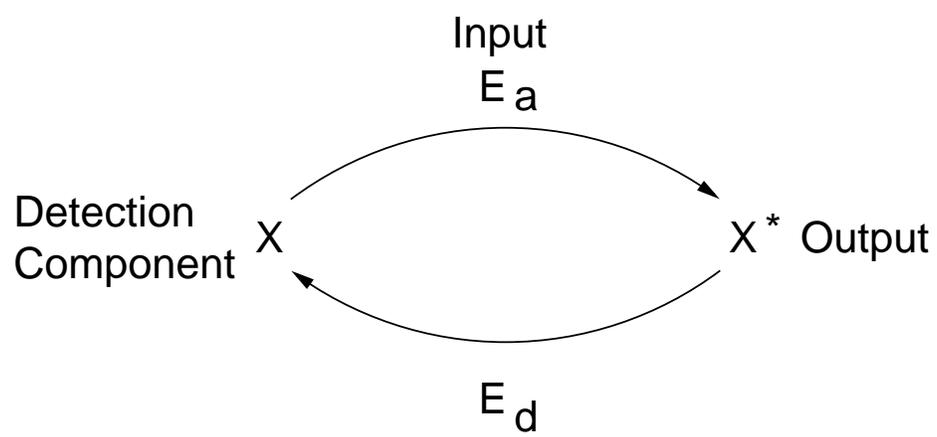}
\caption{Albada and Ten Wolde}
\end{figure}

\newpage
\vspace*{2cm}
\begin{figure}[h]
\centering
\begin{tabular}{cc}
(a)\hspace*{-0.5cm}\epsfig{file=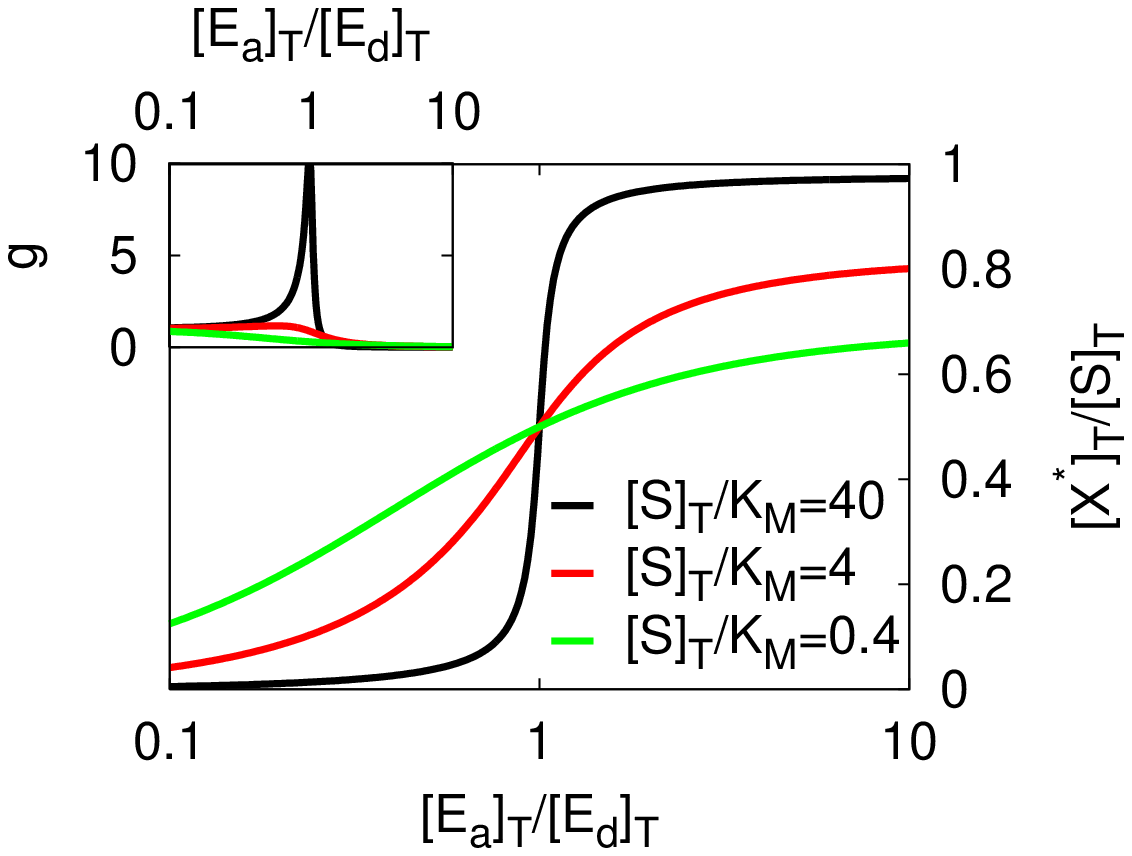,width=0.50\linewidth}&
(b)\hspace*{-0.5cm}\epsfig{file=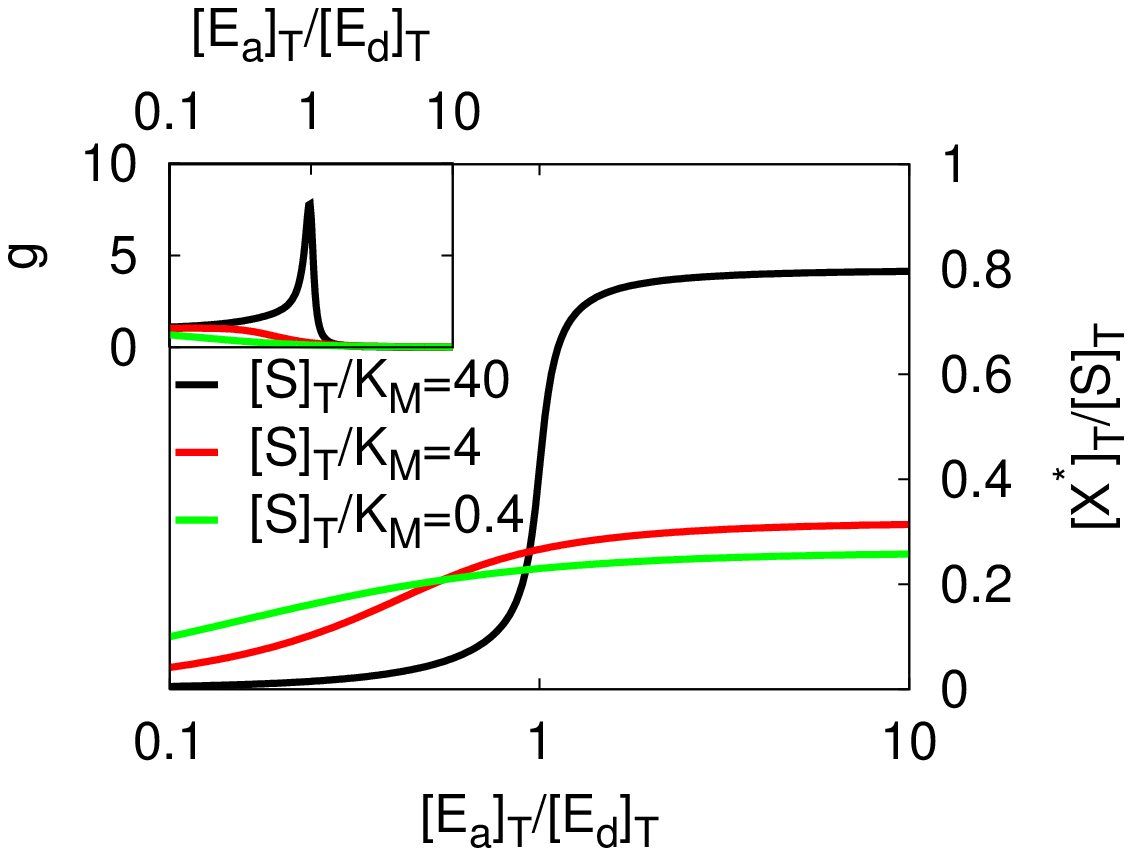,width=0.50\linewidth}
\end{tabular}\caption{Albada and Ten Wolde}
\end{figure}

\newpage
\vspace*{2cm}
\begin{figure}[h]
\centering
\begin{tabular}{cc}
\epsfig{file=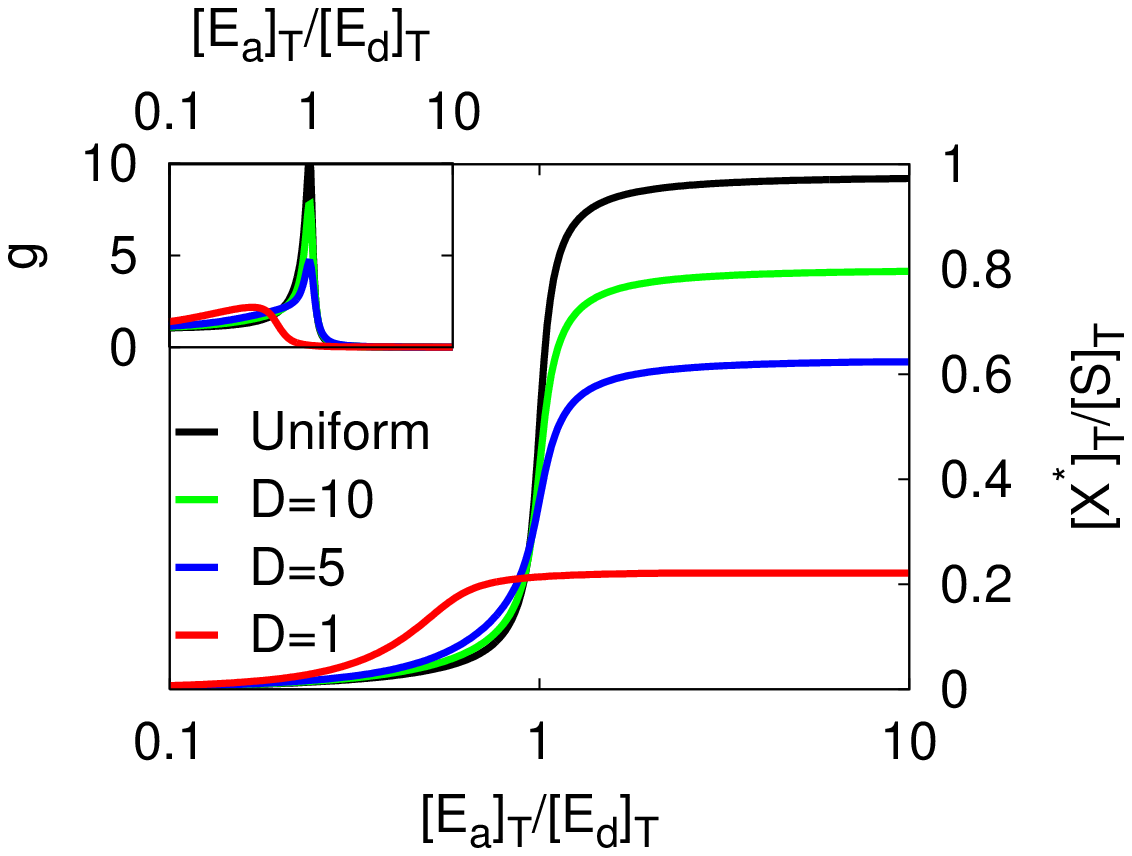,width=0.75\linewidth}
\end{tabular}\caption{Albada and Ten Wolde}
\end{figure}

\newpage
\vspace*{2cm}
\begin{figure}[h]
\centering
\begin{tabular}{cc}
(a)\hspace*{-0.5cm}\epsfig{file=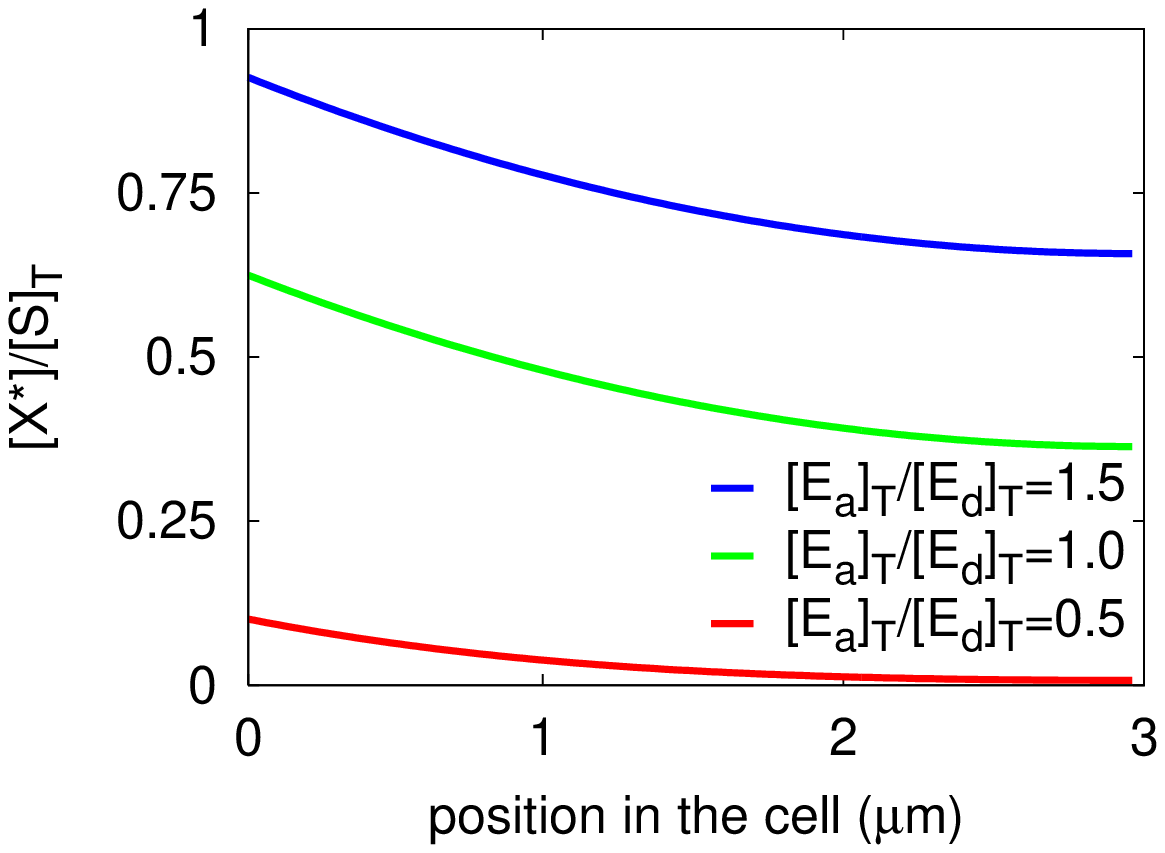,width=0.50\linewidth}&
(b)\hspace*{-0.5cm}\epsfig{file=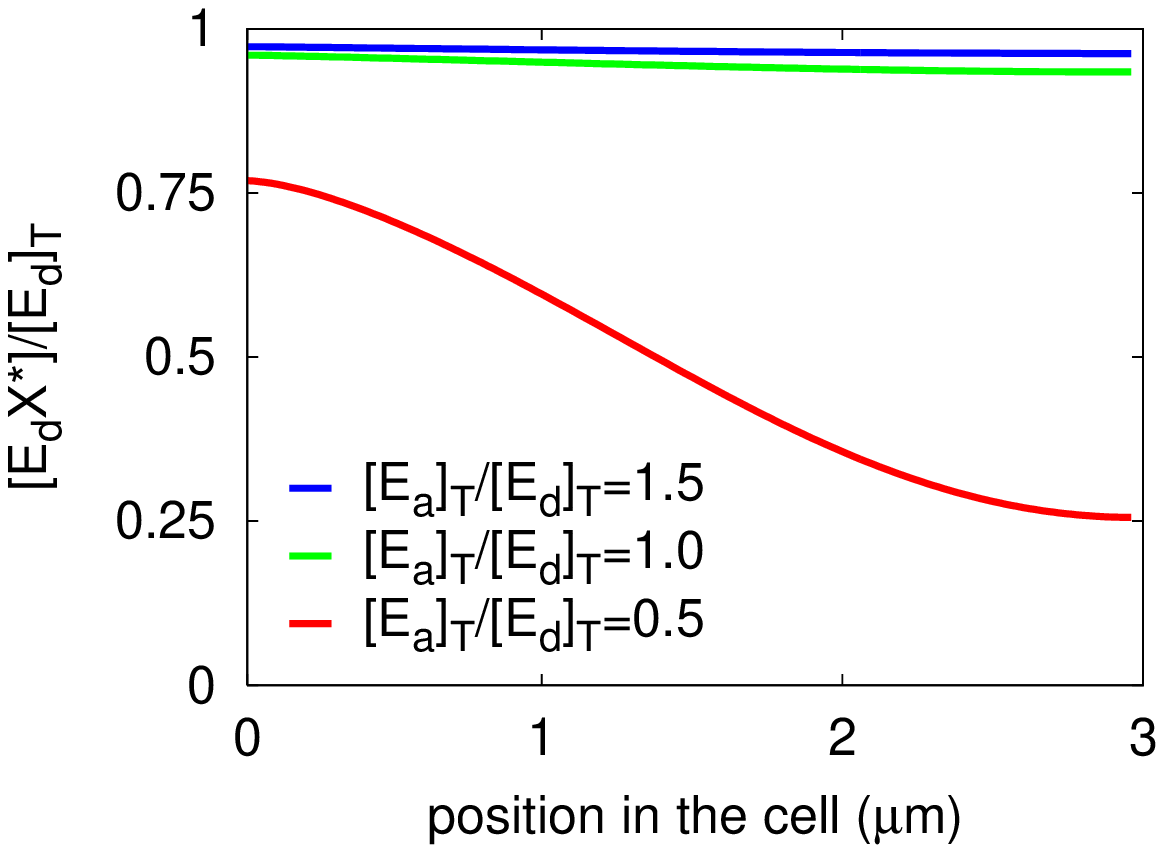,width=0.50\linewidth}
\end{tabular}
\caption{Albada and Ten Wolde}
\end{figure}

\newpage
\vspace*{2cm}
\begin{figure}[h]
\centering
\begin{tabular}{cc}
${\sf \bf [E_a] < [E_d]}$ & ${\sf \bf [E_a] > [E_d]}$\\ 
(a)\epsfig{file=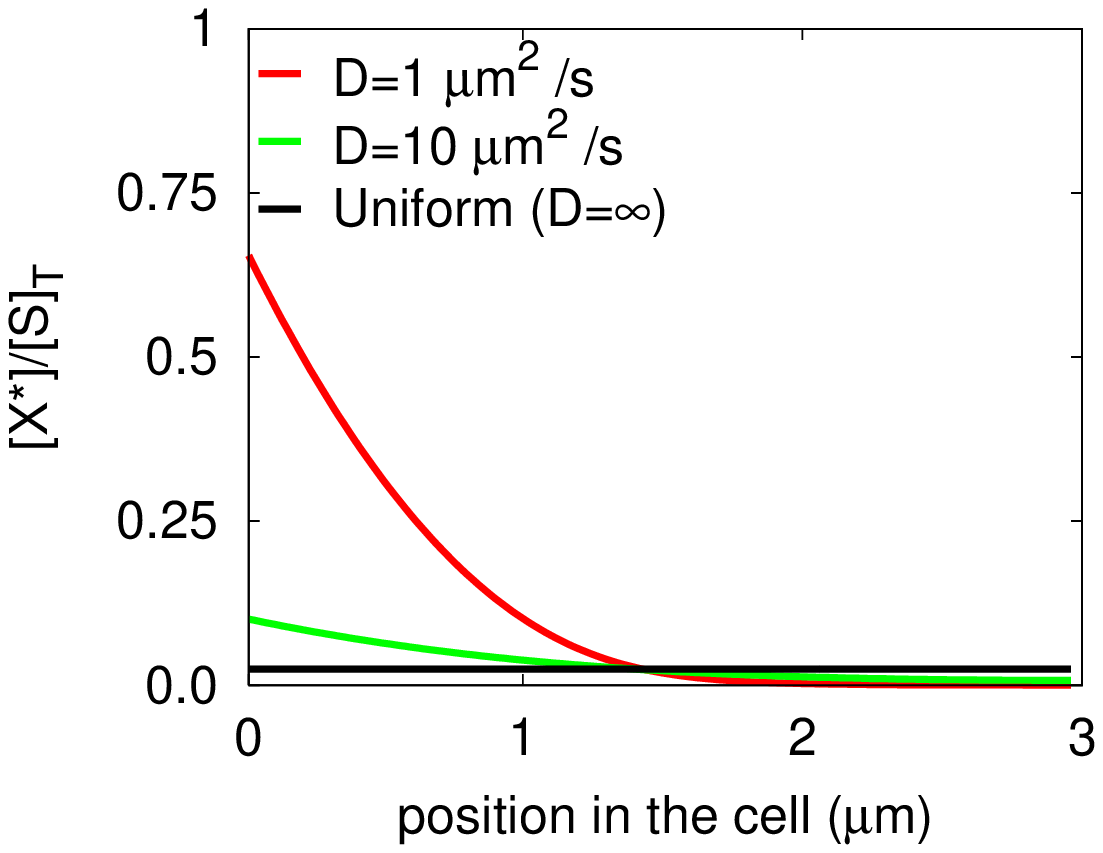,width=0.5\linewidth}&
(d)\epsfig{file=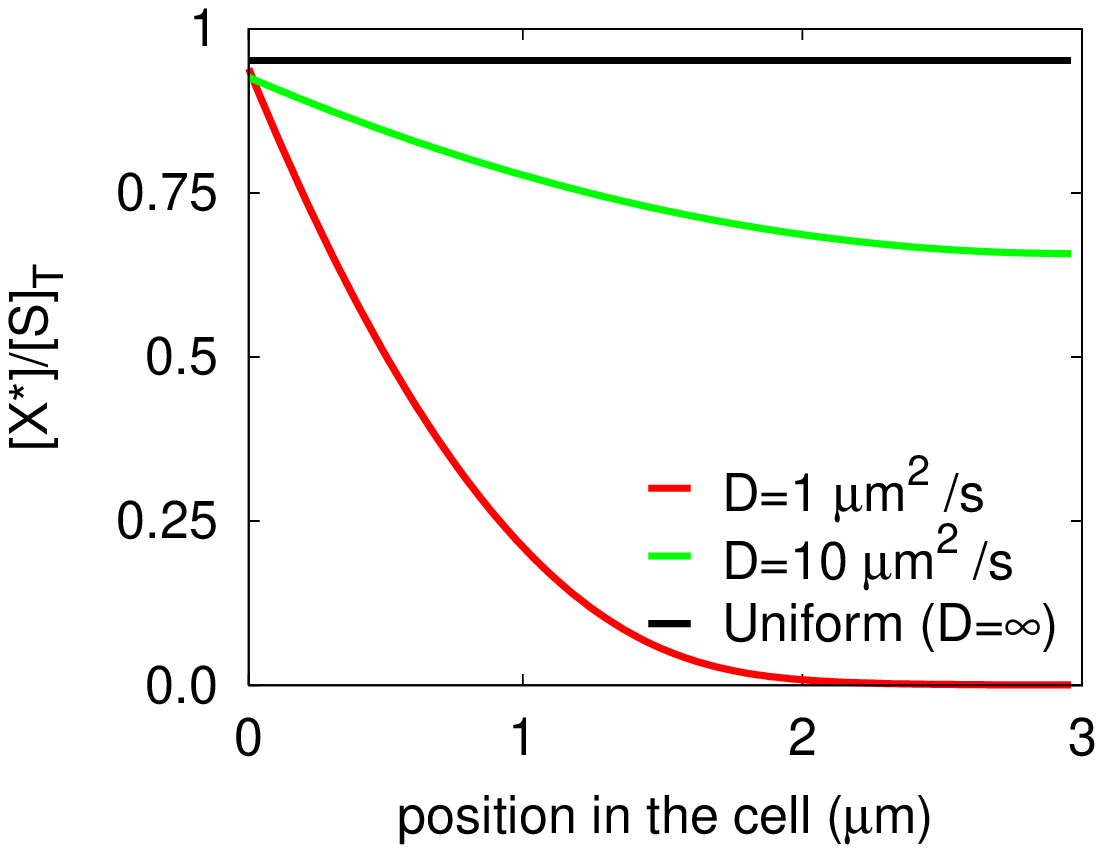,width=0.5\linewidth}\\
(b)\epsfig{file=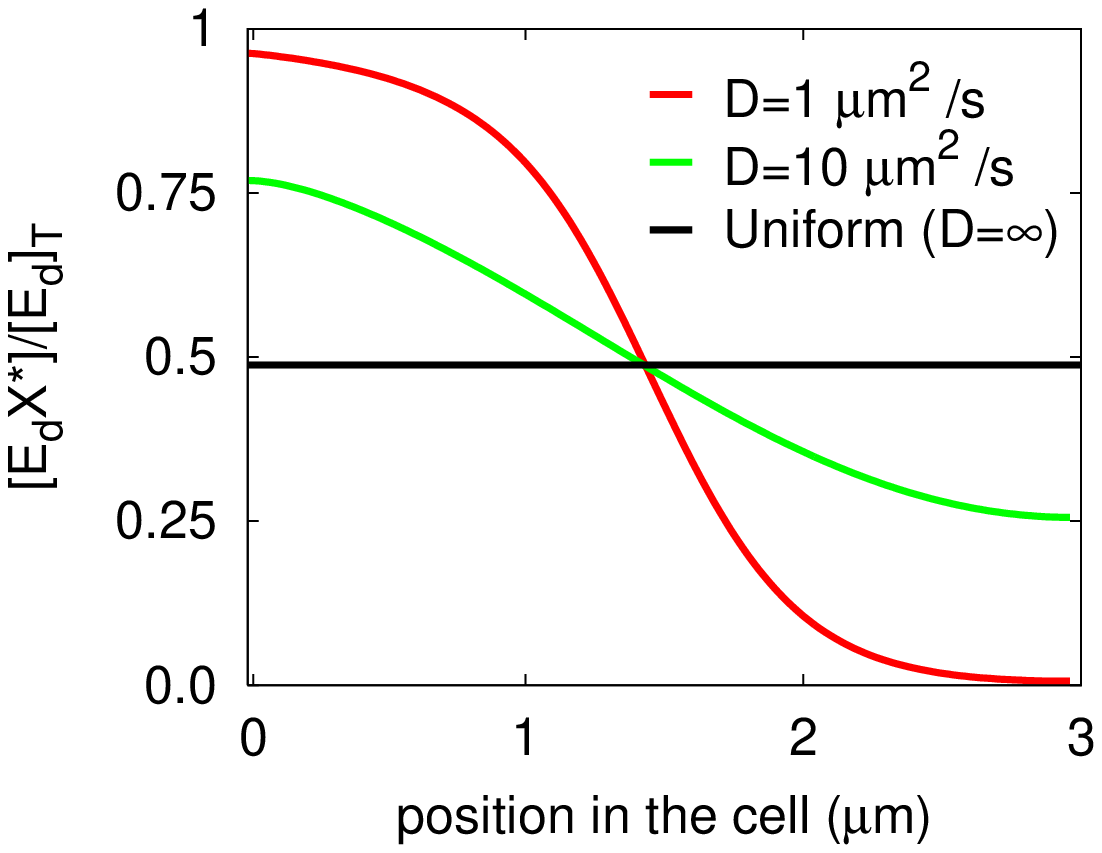,width=0.5\linewidth}&
(e)\epsfig{file=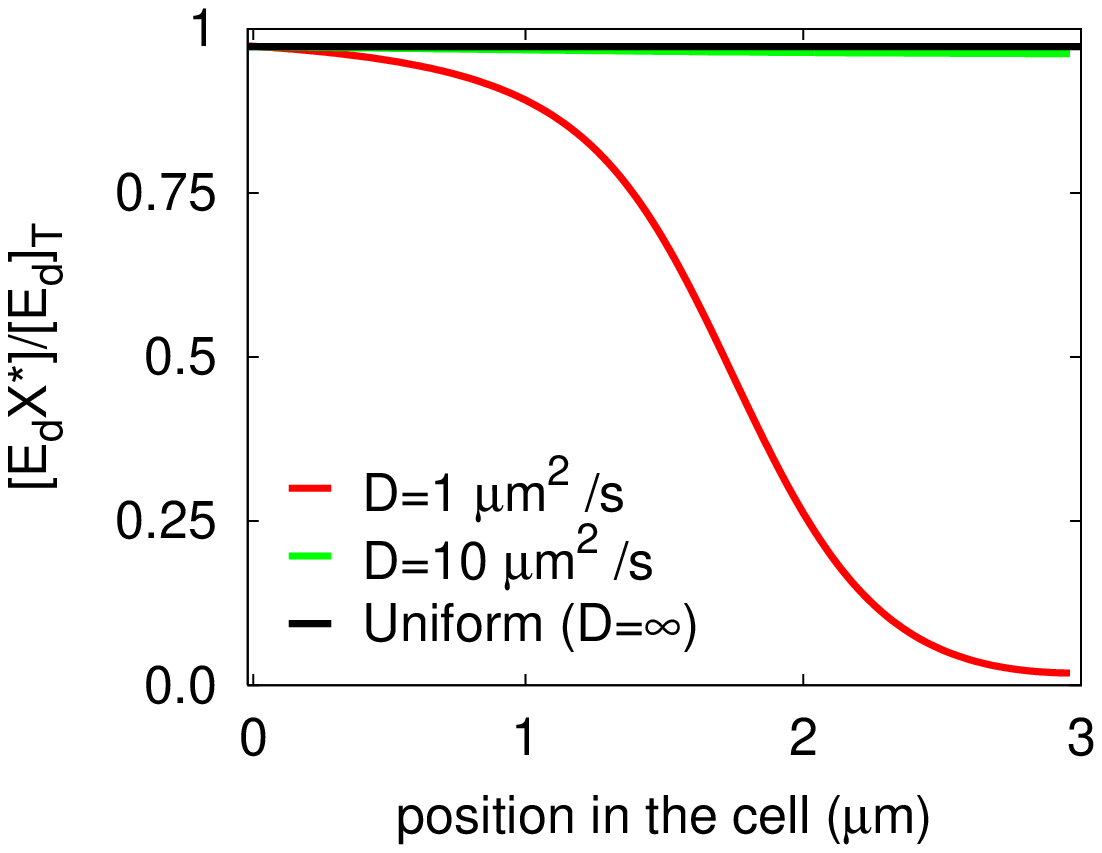,width=0.5\linewidth}\\
(c)\epsfig{file=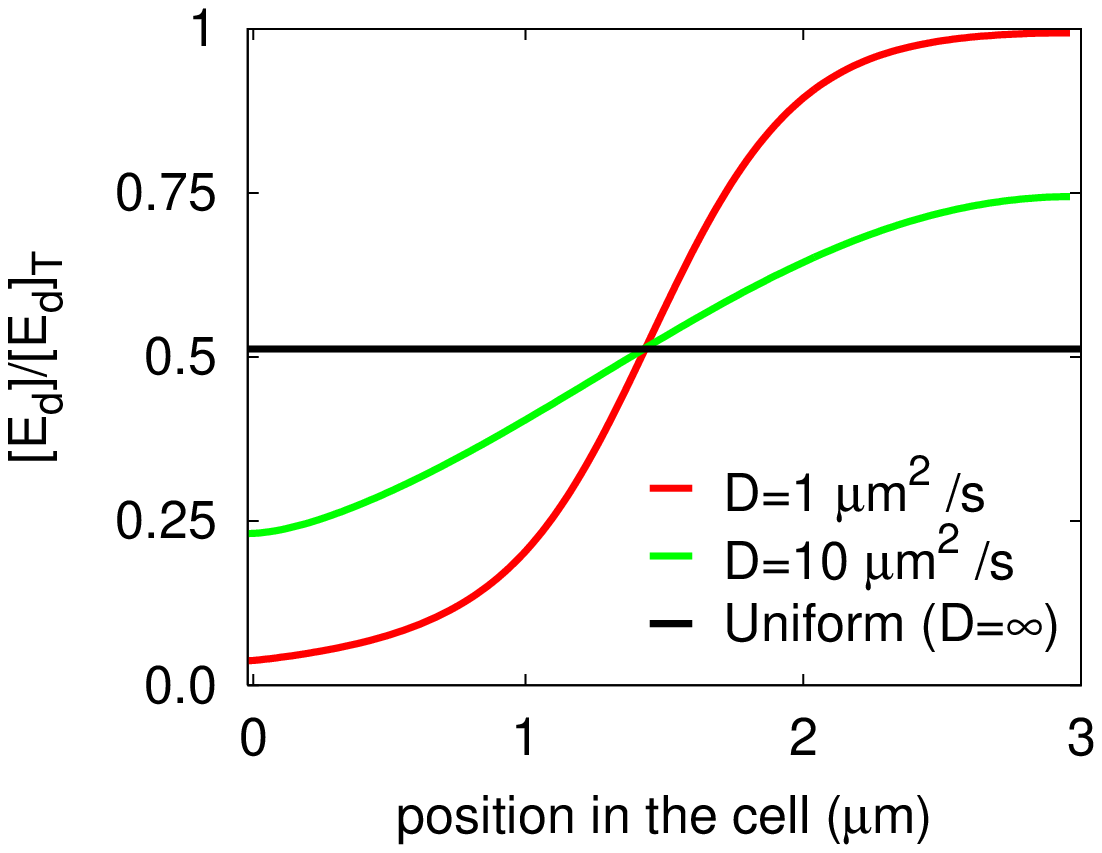,width=0.5\linewidth}&
(f)\epsfig{file=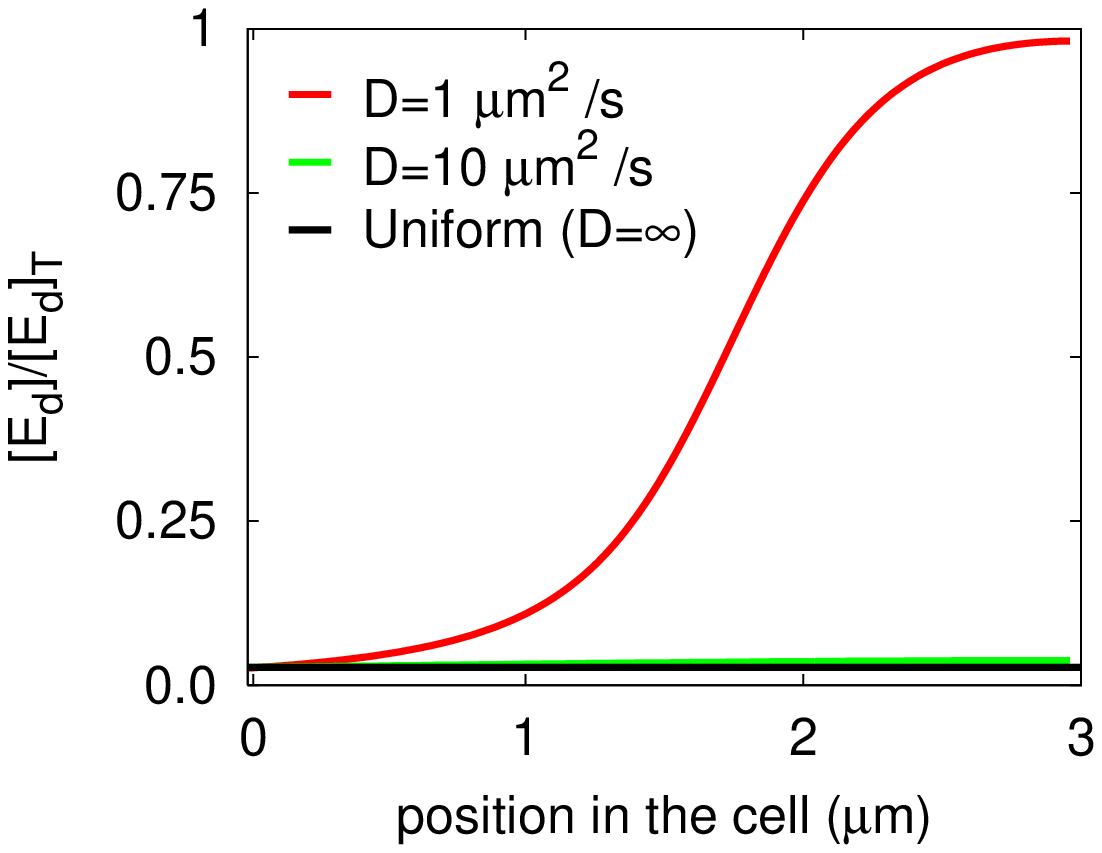,width=0.5\linewidth}\\
\end{tabular}\caption{Albada and Ten Wolde}
\end{figure}

\newpage
\vspace*{2cm}
\begin{figure}[h]
\centering
\begin{tabular}{cc}
(a)\epsfig{file=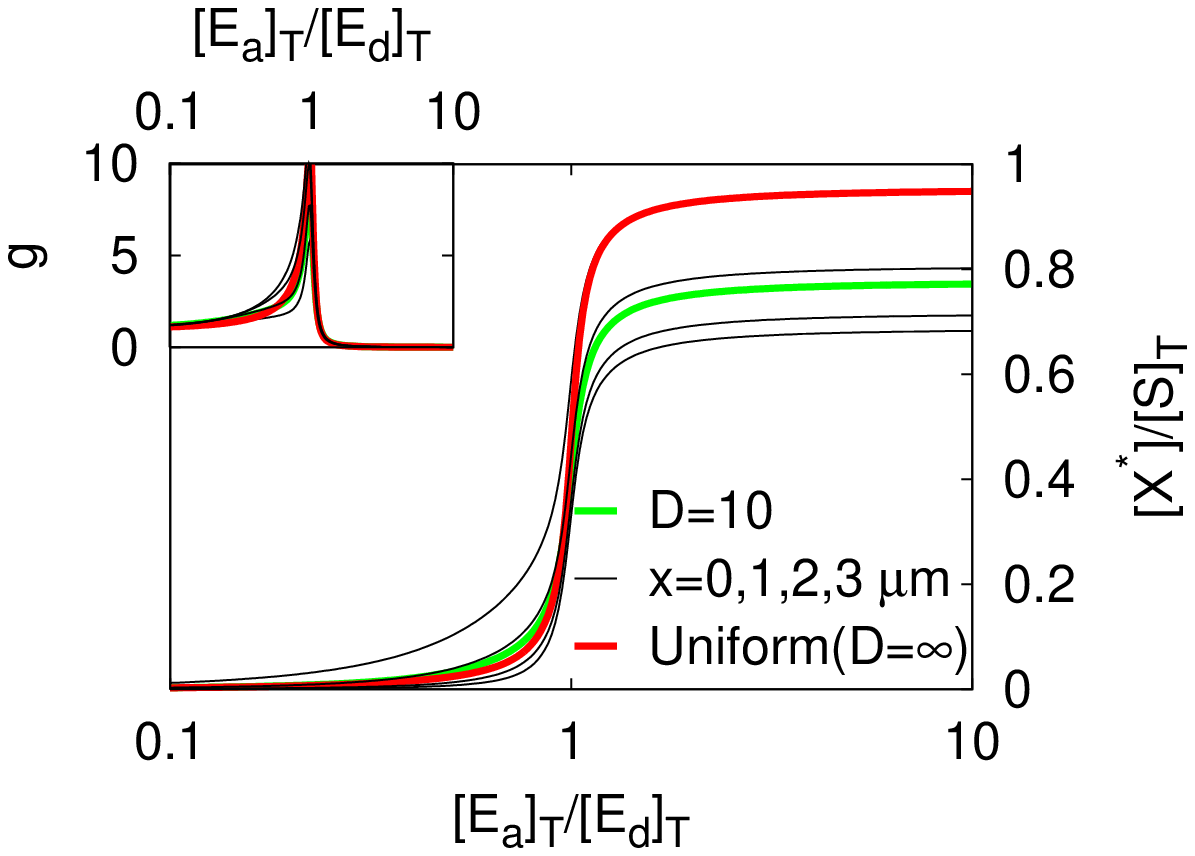,width=0.5\linewidth}&
(b)\epsfig{file=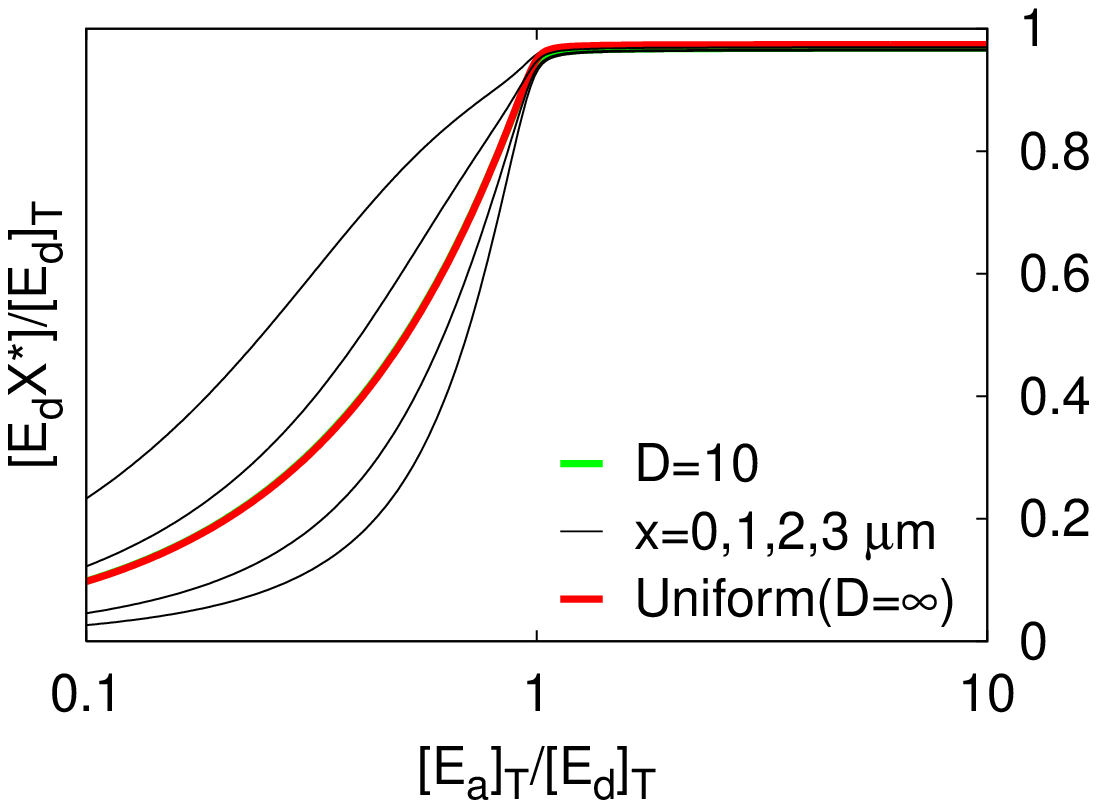,width=0.5\linewidth}\\
\end{tabular}\caption{Albada and Ten Wolde}
\end{figure}

\newpage
\begin{center}
{\large{\bf Text S1}} \\ 
\end{center}

\noindent The spatio-temporal evolution of the concentration profiles
of a push-pull network where the activating enzyme is located at one pole of
the cell, while the deactivating enzyme freely diffuses through the
cytoplasm, is given by Eqs. 9-14 of the main text. To derive the
spatio-temporal evolution of $\cXa$ as given by Eqs. 15-17 of the main
text, we have to make the assumptions that the formation of
the enzyme-substrate complexes is fast and that the diffusion term for
$\cEdXa$ in Eq. 14 can be neglected. The steady-state solutions of
Eqs. 12 and 14 can then be combined with Eq. 9 to yield Eq. 15 of the
main text. We now discuss the steady-state solution of Eq. 15 in the limits of
weak-activation and strong-activation separately.\\[0.2cm]

\noindent {\bf \sf Weak activation} \\ 
In the limit that $\cEaT \ll
\cEdT$, the diffusion term in Eq. 14 of the main text will be small
compared to the other terms in steady-state. Combining the
steady-state solution of that equation with those of Eqs. 9 and 12 of
the main text, yields the following equation for \Xa\ in steady state:
\begin{eqnarray} 
D\frac{{\rm d} ^2 \cXa}{{\rm d} x^2}&=& - k_3 \cEaT \delta (x)\frac{\cX(0)
}{\KMa+\cX(0)}\\
 &&+ k_6 \cEdT \frac{\cXa }{\KMd+\cXa}
\end{eqnarray}
In the limit that $\cEaT \ll \cEdT$, $\cX$ will be large and $\cXa$
will be small. In this limit, the above equation reduces to
\begin{equation}
D\frac{{\rm d} ^2 \cXa}{{\rm d} x^2} \approx -k_3 \cEaT \delta(x) +
k_6/\KMd \cEdT \cXa.
\end{equation}
This equation can be solved with the boundary conditions
\begin{equation}
\label{eq:BC}
-D\left.\frac{\rm d \cXa (x)}{{\rm d} x}\right|_{x=0} = J, \,\,
 \left.\frac{{\rm d}
 \cXa (x)}{{\rm d} x}\right|_{x=L} = 0,
\end{equation}
where $J = k_3\cEaT $ and $L$ is the length of the cell. Defining the effective
deactivation rate of \Xa\ to be $\mu = k_6 / \KMd \cEdT$ and the
characteristic decay length of \Xa\ to be $\lambda = \sqrt{D/\mu}$,
then, if $L \gg \lambda$, the solution of the above equation is
\begin{equation}
\label{eq:cXa_xEalow}
\cXa(x) = \frac{J\lambda}{D}\exp(-x/\lambda).
\end{equation}

This equation predicts that the total amount of \Xa\ in the whole cell
is independent of the diffusion constant $D$. This equation holds only in the
limit that the diffusion term for $\cEdXa$ in Eq. 14 of the main text
can be neglected and if the deactivating enzyme \Ed\ is unsaturated,
meaning that the total deactivation rate $\gamma$ is proportional to
the concentration of \Xa. These assumptions are only accurate when
$\cEaT \ll \cEdT$. 

As $\cEaT$ increases, $\cXa$ increases, and close to the pole where
the activating enzyme resides, the deactivating enzyme does become
saturated with \Xa. This lowers the deactivation rate {\em per
particle}, which in turn increases the concentration of \Xa\ with
respect to that in a system where both enzymes are uniformly
distributed in space.

The significance of the diffusion term for $\cEdXa$ impedes an
analytical derivation of $\cXa(x)$. However, when $\cEaT <
\cEdT$ and $\cX$ is large, $\cEaX$ is essentially constant and
independent of the diffusion constant. Under this condition we can
prove that the total amount of \Xa\ must decrease with increasing
diffusion constant, implying that in this regime the uniform network
will actually respond weaker than the spatially non-uniform
network. Integrating Eq. 14 over the length of the cell
reveals that in steady state $k_4 \int_0^L dx \cEd(x) \cXa(x) =
(k_5+k_6) \int_0^L dx \cEdXa(x)$. Combining Eq. 9 and
Eq. 14 reveals that in steady state $k_3 \cEaX = k_6 \int_0^L
dx \cEdXa (x)$. Hence, in steady state $\cEaX = c \int_0^L dx \cEd(x)
\cXa(x)$, where $c$ is a constant. This means that if $\cEaX$ is
independent of the diffusion constant, also $\int_0^L dx \cEd(x)
\cXa(x)$ must be independent of the diffusion constant. Now, when the
diffusion constant decreases, the profiles $\cEd(x)$ and $\cXa(x)$
vary more strongly in space and become less overlapping, as
Figs. 5a and 5c show. The decrease in
overlap between $\cEa(x)$ and $\cXa(x)$ would tend to decrease
$\int_0^L dx \cEd (x) \cXa (x)$. To compensate for this, the total amount of \Xa\ must
increase with decreasing diffusion constant, when $\cEaT < \cEdT$ and
$\cEaX$ is constant.
\\[0.2cm]

\noindent{\bf \sf Strong activation}\\
Combining Eq. 9 with
Eq. 14 yields, in steady state:
\begin{equation}
\label{eq:X_Eahigh}
 D\frac{{\rm d}^2 \cXa (x)}{{\rm d} x^2} = - k_3 \cEaX \delta (x)- D
 \frac{{\rm d}^2 \cEdXa (x)}{{\rm d} x^2}  + k_6 \cEdXa (x) .
\end{equation}
This equation for the steady-state concentration profile of \Xa\ is
exact. Nevertheless, the diffusion term for $\cEdXa$ impedes a
transparant analytical solution. However, if $\cEaT > \cEdT$ and $D$
is not too low, then essentially all of the deactivating enzyme is
saturated, and $\cEdXa (x) = \cEdT (x) \equiv \cEd (x) + \cEdXa (x)$
(see Fig.5e of main text). Moreover, combining Eqs. 13
and 14 of the main text reveals that $\cEdT(x)$ is constant in space,
if, as assumed here, the diffusion constants of the enzyme \Ed, and
that of the enzyme bound to its substrate, \EdXa, are the same.
 Hence, in the
limit that $\cEa$ is high and \Ed\ is saturated, the second term on
the right-hand-side of the above equation is zero, the third
term is constant, and the equation reduces to
\begin{equation}
\label{eq:XaEahsim}
D\frac{{\rm d}^2 \cXa (x)}{{\rm d} x^2} = - k_3 \cEaX \delta (x)  + k_6 \cEdT.
\end{equation}
This equation can be solved with the boundary conditions in
Eq.~\ref{eq:BC} with $J=k_3\cEaX$.  The solution
is
\begin{equation}
\label{eq:cXa_x_Eahigh}
\cXa(x) = c_0 + c_1 x + \frac{1}{2} c_2 x^2,
\end{equation}
where $c_2 = k_6 \cEdT / D$, obtained from the solution of
Eq.~\ref{eq:XaEahsim} in the domain $0<x<L$, and $c_1 = -k_6 \cEdT L /
D$, obtained from the boundary condition at $x=L$. The coefficient
$c_0 = \cXa(0)$ can be obtained from the boundary condition at $x=0$:
$k_6\cEdT L = k_3 \cEaX$. First, we note that the total substrate
concentration in the {\em cytoplasm}, ${\rm [S]_c}(x) \equiv \cX(x) +
\cXa(x) + \cEdXa(x) = {\rm [S]_c}$, is constant in space, because the
diffusion constants of all the diffusing components are equal. Hence,
the total substrate concentration in the whole cell is $\cST \equiv
{\rm [S]_c} + \cEaX/L = \cST = \cX(x) + \cEaX/L + \cXa(x) + \cEdT$
(where we have used that $\cEdXa(x) \approx \cEdT$). The
concentration $\cX(0)$ can be obtained by solving Eq. 12 of the main text
in steady state, which gives $\cX(0) = \KMa \cEaX / (L \cEaT -
\cEaX)$. Combining these expressions with the boundary condition at
$x=0$, and using that when $\cEaT$ is high, $\cEaT \gg \cEaX$, yields
$c_0 = \cXa(0) = \cST - \cEdT(1+k_6/k_3(1+\KMa/\cEaT))$.

\end{document}